\documentclass[12pt]{article}

\newsavebox{\ns}
\newsavebox{\dbrane}
\newsavebox{\dbshort}

\usepackage{epsfig}
\usepackage{amssymb}

\def\appendix{{\newpage\section*{Appendix}}\let\appendix\section%
        {\setcounter{section}{0}
        \gdef\thesection{\Alph{section}}}\section}

\def\be{\begin{eqnarray}}
\def\ee{\end{eqnarray}}

\newcommand{\nn}{\nonumber}
\newcommand\para{\paragraph{}}
\newcommand{\ft}[2]{{\textstyle\frac{#1}{#2}}}
\newcommand{\eqn}[1]{(\ref{#1})}

\newcommand\bomega{\mbox{\boldmath $\omega$}}

\def\Dslash{\,\,{\raise.15ex\hbox{/}\mkern-12mu D}}
\def\Dbarslash{\,\,{\raise.15ex\hbox{/}\mkern-12mu {\bar D}}}
\def\delslash{\,\,{\raise.15ex\hbox{/}\mkern-9mu \partial}}
\def\delbarslash{\,\,{\raise.15ex\hbox{/}\mkern-9mu {\bar\partial}}}
\def\pslash{\,\,{\raise.15ex\hbox{/}\mkern-9mu p}}
\def\calDslash{\,\,{\raise.15ex\hbox{/}\mkern-12mu {\cal D}}}

\newcommand\vkn{{\cal V}_{k,N}}
\newcommand\ikn{{\cal I}_{k,N}}
\newcommand\tvkn{{\tilde{\cal V}}_{k,N}}

\newcommand\mkn{{\cal M}_{k,N}}

\begin{document}
\pagestyle{plain}
\setcounter{page}{1}
\newcounter{bean}
\baselineskip16pt

\begin{titlepage}

\begin{center}
\today
\hfill hep-th/0306150\\
\hfill MIT-CTP-3388 \\

\vskip 1.5 cm
{\large \bf Vortices, Instantons and Branes}
\vskip 1 cm 
{Amihay Hanany and David Tong}\\
\vskip 1cm
{\sl Center for Theoretical Physics, 
Massachusetts Institute of Technology, \\ Cambridge, MA 02139, U.S.A.\\
{\tt hanany,dtong@mit.edu}\\}

\end{center}

\vskip 0.5 cm
\begin{abstract}
The purpose of this paper is to describe a relationship between the 
moduli space of vortices and the moduli space of instantons. 
We study charge $k$ vortices in $U(N)$ Yang-Mills-Higgs 
theories and show that the moduli space is  isomorphic to a 
special Lagrangian submanifold of the moduli space of $k$ 
instantons in non-commutative $U(N)$ Yang-Mills theories. This 
submanifold is the fixed point set of a $U(1)$ action on the instanton 
moduli space which rotates the instantons in a plane. To derive this 
relationship, we present a D-brane construction in which the dynamics 
of vortices is described by the Higgs branch of a $U(k)$ gauge theory with 
4 supercharges which is a 
truncation of the familiar ADHM gauge theory. 
We further describe a moduli space construction for semi-local 
vortices, lumps in the ${\bf CP}^N$ and Grassmannian sigma-models, 
and vortices on the non-commutative plane.
We argue that this relationship between vortices and instantons underlies many of 
the quantitative similarities between quantum field theories in two and 
four dimensions.

\end{abstract}

\end{titlepage}

\section{Introduction and Preview}

The moduli space of a supersymmetric system is defined as the set 
of classically massless, or light, degrees of freedom. The beauty of 
this concept lies in the fact that much of the low-energy behaviour of 
the system  
may  be encoded as geometrical features on the moduli space. 
Whether the subject be string compactifications, the dynamics of 
gauge theories, or the interactions of solitons, the moduli space 
approximation provides an effective, and tractable, approach to 
extract the infra-red quantum properties of the system. 
\para
One particularly useful geometric feature of the moduli space is the 
metric, describing the kinetic interactions of the system. Our 
interest in this paper will be focused on the moduli space of 
solitons, specifically vortices. In this case, the relevance of the metric was first 
revealed by Manton who showed that geodesics on the  
moduli space track the classical scattering of solitons \cite{manton}. 
\para
It is common lore that for dynamics exhibiting 8 or more supercharges, 
the metric on the moduli space is exactly calculable. For 
theories with 4 supercharges or less, the metric can, in general, only be computed 
in asymptotic regimes. In the context of solitons, both Yang-Mills instantons and 
monopoles preserve up to 8 supercharges and indeed exact, albeit somewhat 
implicit, expressions for the metrics are known using the techniques of 
\cite{adhm,nahm}\footnote{For particularly simple cases, more explicit descriptions 
also exist. See, for example \cite{ah}.}. 
In contrast, vortices preserve a maximum of only 4 supercharges, and 
knowledge of the metric is currently restricted to the situation 
where the solitons are well-separated \cite{samols,manspe}.
\para
Nevertheless, it is possible to make progress in supersymmetric quantum 
field theories  
even when the moduli space metric is not known. This is because, as first 
emphasised by Witten \cite{witten81}, 
many of the simplest quantities of interest depend only on topological 
characteristics of the moduli space. For example, the supersymmetric 
bound states of solitons are related to various cohomology classes of 
the moduli space 
\cite{witten81,boundstate}\footnote{Questions of $L^2$ normalisability 
mean that asymptotic behaviour of the metric is also required.}. 
Similarly non-perturbative contributions to 
BPS correlation functions, which involve integrals over the moduli space 
of instantons, often reduce to topological invariants 
\cite{instanton,instanton2}. Thus, for many purposes it suffices to know 
only crude topological information about the moduli space. 
\para
The purpose of this paper is to describe the moduli space of vortices in 
$U(N)$ Yang-Mills-Higgs theories where the gauge group is broken 
completely by $N$ fundamental scalar fields. 
The theory has a mass gap and exhibits vortices, labeled 
by the winding number $k$ of the magnetic field,
\be
{\rm Tr}\ \int B = -2\pi k
\label{first}\ee
Here we summarise our main results. 
We start in Section 2 with a study of the moduli space of charge $k>0$ vortices 
which we shall denote as ${\cal V}_{k,N}$. Our first result concerns the 
real dimension of the moduli space which, using index theory techniques, we 
show to be
\be
{\rm dim}({\cal V}_{k,N})=2kN
\label{dim}\ee
In Section 3 we present a brane construction of the vortices, from which we 
extract a description of $\vkn$ as a $U(k)$ symplectic 
quotient of ${\bf C}^{k(N+k)}$. 
This quotient construction is most easily described as the Higgs branch 
of a $U(k)$ gauge theory with four supercharges, coupled to a single 
adjoint chiral multiplet and $N$ fundamental chiral multiplets. 
\para
The moduli space $\vkn$ naturally inherits a metric from the K\"ahler 
quotient construction. This does not agree with the Manton metric 
describing the classical scattering of solitons. Given our 
discussion above, this is neither unexpected nor an obstacle to 
utilising our construction for further calculations. As we shall 
see, the inherited metric is a deformation of the Manton 
metric, preserving the K\"ahler property, the isometries and 
the asymptotic form. 
\para
The parametric scaling of the dimension \eqn{dim} is reminiscent of the 
moduli space of $k$ instantons in a $U(N)$ gauge theory, which we shall 
denote as $\ikn$. Recall that the real dimension of the instanton moduli 
space is
\be
{\rm dim}(\ikn)=4kN
\nn\ee
Moreover, those familiar with instanton moduli spaces will have recognised 
the quotient construction of $\vkn$ as a truncated version of the 
ADHM quotient \cite{adhm}. In Section 4, we make this 
relationship more explicit and show that the moduli space of vortices $\vkn$ 
is a complex middle-dimensional submanifold (or, since $\ikn$ is hyperK\"ahler, 
equivalently a special Lagrangian submanifold) of the resolved  instanton moduli space  
$\ikn$. We further show that $\vkn$ may be realised as the fixed point set of a 
holomorphic $U(1)$ action on $\ikn$, descending from the rotations of 
instantons in a plane. 
%\para
%The correspondence between the vortex and instanton moduli spaces exists only 
%when the singularities of the latter are resolved in the manner proscribed 
%by Nakajima \cite{nak}. As is well 
%known, a physical manifestation of this resolution was provided by Nekrasov and 
%Schwarz who showed that it occurs for instantons 
%on non-commutative ${\bf R}^4$ \cite{ns}. Writing the non-commutative deformation 
%parameter as $\theta$, we shall see that this is related to the gauge 
%coupling $e^2$ of the theory containing vortices,
%\be
%\frac{2\pi}{e^2}=\theta.
%\nn\ee
\para
In Section 5, we generalise this construction to vortices in $U(N)$ gauge 
theories with $N_f=N+M$ flavours. For abelian gauge theories, such objects 
have been well studied and are known as semi-local vortices. In the 
strong coupling limit these vortices become lump 
solutions on the ${\bf CP}^{M}$ Higgs branch of the theory. For the 
non-abelian theory, these vortices are related to lumps in the 
$G(N,N_f)$ Grassmannian sigma-model of $N$ planes in ${\bf C}^{N_f}$. 
We denote these moduli spaces of vortices as
$\hat{\cal V}_{k,(N,M)}$ (note that $\vkn\cong \hat{\cal V}_{k,(N,0)}$). 
The dimension of the moduli space is,
\be
{\rm dim}(\hat{\cal V}_{k,(N,M)})=2k(N+M).
\nn\ee
We again give a brane construction as well as a quotient construction of the moduli space and 
explain how it can 
be described as the fixed point set of a (different) holomorphic action on 
the moduli space of instantons ${\cal I}_{k,N_f}$.
\para
In Section 6, we consider the Yang-Mills-Higgs theory defined on the spatial non-commutative 
plane with $[x^1,x^2]=-i\vartheta$. 
We describe how the moduli space of vortices changes as $\vartheta$ is varied. 
We show that the moduli spaces may become singular, cease to exist, or 
undergo interesting topology changing transitions for different values of $\vartheta$. 
We end in Section 7 with conclusions and a discussion.

\section{Vortices}

Our starting point is the maximally supersymmetric theory admitting 
vortex solutions which, for concreteness, we choose to live in 
$d=2+1$ dimensions with ${\cal N}=4$ supersymmetry\footnote{Supersymmetric 
theories which admit vortices exist in any 
dimension between 1+1 and 5+1. The discussion of quantum effects,  
particular to each case, is very interesting but will be left for future work.}. 
Our theory includes a $U(N)$ vector multiplet, consisting of a gauge field $A_\mu$, 
a triplet of adjoint scalar fields $\phi^r$, $r=1,2,3$ and their fermionic partners. 
To these we couple $N$ fundamental 
hypermultiplets, each of which contains two complex scalars 
$q$ and $\tilde{q}$, and their partner fermions. 
As well as the $U(N)_G$ gauge symmetry, the Lagrangian also 
enjoys a $SU(N)_F$ flavour symmetry. Under these two groups, the $q$ field 
transforms as $({\bf N},\bar{\bf N})$, while $\tilde{q}$ transforms as 
$(\bar{\bf N},{\bf N})$. In the following we take both $q$ and $\tilde{q}$ 
to represent $N\times N$ matrices,
\be
q=q^a_{\ i}\ \ ,\ \ \tilde{q}=\tilde{q}^i_{\ a}\ \ \ \ \ \ \ \ a,i=1,\ldots,N
\nn\ee
where the $a$ index furnishes a representation under $U(N)_G$ while the 
$i$ index refers to $SU(N)_F$. In this notation, 
the bosonic part of the Lagrangian reads\footnote{Our conventions: we choose a 
Hermitian connection with $F_{\mu\nu}=\partial_{\mu}A_\nu-\partial_\nu A_\mu-
i[A_\mu,A_\nu]$ and ${\cal D}_\mu q =\partial_\mu q -iA_\mu q$. All gauge and 
flavour indices are implicit and assumed summed, with the exception of $r,s=1,2,3$ 
which is explicit and summed.}, 
\be
{\cal L}&=&-{\rm Tr}\left[\frac{1}{4e^2}F_{\mu\nu}F^{\mu\nu}+\frac{1}{2e^2}
{\cal D}\phi^r{\cal D}\phi^r+
{\cal D}_\mu q^\dagger
{\cal D}^\mu q+{\cal D}_\mu\tilde{q}{\cal D}^\mu\tilde{q}^\dagger
+e^2|q\tilde{q}|^2\right.\nn\\ 
&&\ \ \ \ \ \ \left.+\frac{1}{2e^2}[\phi^r,\phi^s]^2
+(\tilde{q}^\dagger\tilde{q}+
qq^\dagger)\phi^r\phi^r
+\frac{e^2}{2}\left(
qq^\dagger-\tilde{q}^\dagger\tilde{q}-\zeta{\bf 1}_N\right)^2\right]
\label{lag}\ee
The final term in the Lagrangian is a D-term and  
includes a Fayet-Iliopoulos parameter $\zeta$, 
which we take to be strictly positive $\zeta>0$. The presence of this  
parameter induces symmetry breaking with the unique vacuum, up 
to Weyl permutations, given by
\be
q^a_{\ i}=\sqrt{\zeta}\delta^a_{\ i}\ \ \ \ \ ,\ \ \ \ \ \ \ 
\tilde{q}^i_{\ a}=0\ \ \ \ \ ,\ \ \ \ \ \ \ \phi^r=0
\nn\ee
The ground state of the theory is a gapped, colour-flavour locking 
phase with the symmetry breaking pattern
\be
U(N)_G\times SU(N)_F\rightarrow SU(N)_{\rm diag}
\nn\ee
The breaking of the overall $U(1)_G$ gauge symmetry ensures the existence of 
vortex solutions in the theory. These vortices obey a Bogomoln'yi 
bound which is the natural generalisation of the usual abelian vortex 
bound \cite{bog} and may be simply determined by the standard trick of completing the 
square in the Hamiltonian. It will turn out that the most general vortex 
solutions involve only the fields $q$ and $B=F_{12}$, and we choose to set 
the remaining fields to zero at this stage. Restricting to time independent 
configurations, the Hamiltonian reads, 
\be
{\cal H}&=&{\rm Tr}\left[\frac{1}{2e^2}B^2+|{\cal D}_1q|^2+|{\cal D}_2q|^2 
+\frac{e^2}{2}(qq^\dagger - \zeta{\bf 1})\right]\nn\\
&=&{\rm Tr}\left[\frac{1}{2e^2}(B\mp e^2(qq^\dagger-\zeta{\bf 1}))^2
+|{\cal D}_1q^\dagger\pm{\cal D}_2q^\dagger|^2\mp\zeta B\right]\nn\\
&\geq& 2\pi\zeta|k|
\nn\ee
where $k\in{\mathbb Z}$ is the winding number of the configuration  
defined in \eqn{first}. Choosing $k>0$, the bound is saturated by 
configurations satisfying the first order Bogomoln'yi equations 
which, for once, we write with all indices explicit to emphasise their 
matrix nature
\be
B^a_{\ b}&=&e^2(q^a_{\ i}q^{i\dagger}_{\ b}-
\zeta\delta^a_{\ b}) \nn\\
{\cal D}_zq^a_{\ i}&=&\partial_zq^a_{\ i}-i(A_z)^a_{\ b}q^b_{\ i}=0
\label{bog}\ee
where we have introduced the complex coordinate on the spatial plane 
$z=x^1+ix^2$. 
\para
The main purpose of this paper will be to study the moduli 
space of solutions to these equations. We denote the moduli space of 
charge $k$ vortices in the $U(N)$ Yang-Mills-Higgs theory as $\vkn$. 
We start with a study of the linearised equations to determine the 
dimension of $\vkn$. The reader uninterested in the details of the 
index theorem may skip to the following subsection where basic properties of 
$\vkn$ are discussed, taking with them the following punchline:
\be
{\rm dim}(\vkn)=2kN
\label{dimvkn}\ee

\subsubsection*{An Index Theorem}

In this section, we prove the result \eqn{dimvkn} by studying 
the fluctuations $(\dot{A},\dot{q})$ around a given solution. Our method follows 
closely the work of E. Weinberg \cite{erick} who analysed the moduli space in the 
abelian case $N=1$. The linearised Bogomoln'yi matrix equations are
\be
{\cal D}^a_z\dot{A}_{\bar{z}}-{\cal D}_{\bar{z}}^a\dot{A}_z&=&
\frac{ie^2}{2}(\dot{q}q^\dagger+q\dot{q}^\dagger)\nn\\
{\cal D}_z^f\dot{q}&=&i\dot{A}_zq
\label{linbog1}\ee
and are to be augmented with a gauge fixing condition, for which we choose Gauss' law, 
\be
{\cal D}^a\dot{A}_{\bar{z}}+{\cal D}^a_{\bar{z}}\dot{A}_z&=& 
-\frac{ie^2}{2}(\dot{q}q^\dagger-q\dot{q}^\dagger)
\label{gauge}\ee
which can be combined with the first of the equations in \eqn{linbog1} to 
give
\be
{\cal D}^a_{\bar{z}}\dot{A}_z&=&-\frac{ie^2}{2}\dot{q}q^\dagger
\label{linbog2}\ee
The observant reader will have noticed the appearance of superscripts on 
the covariant derivatives, which are there to remind us of the $U(N)_G$ 
representation of the field on which they act:
\be
{\cal D}^aX=\partial X-i[A,X]\ \ \ ,\ \ \ 
{\cal D}^fY=\partial X - iAY\ \ \ ,\ \ \ 
{\cal D}^{\bar{f}}Y^\dagger=\partial Y^\dagger+iY^\dagger A
\nn\ee
Before proceeding, notice that it is possible to rescale the gauge field 
$A\rightarrow A/e$ and coordinate $z\rightarrow ez$ to remove $e^2$ from the 
equations. The number of zero modes is therefore independent of $e^2$ and we 
use this freedom to set $e^2=2$ which simplifies the linearised Bogomoln'yi 
equations somewhat so they can be written as,
\be
\Delta\eta\equiv\left(\begin{array}{cc}i{\cal D}^a_{\bar{z}} & -q^{r\dagger} \nn\\
q^r & i{\cal D}_z^f \end{array}\right)\left(\begin{array}{c} \dot{A}_z \nn\\ \dot{q} 
\end{array}\right)=0
\nn\ee
where the superscript in $q^r$ denotes the fact that the matrix $q$ 
acts as right multiplication. 
We now define the index of $\Delta$ as
\be
{\cal J}=\lim_{M^2\rightarrow 0} {\cal J}(M^2)\equiv\lim_{M^2\rightarrow 0}
\left[ {\rm Tr}
\left(\frac{M^2}{\Delta^\dagger\Delta+M^2}\right)-{\rm Tr}\left(
\frac{M^2}{\Delta\Delta^\dagger+M^2}\right)\right]
\nn\ee
which counts the number of complex zero modes of $\Delta$ minus the number of 
zero modes of $\Delta^\dagger$. 
Let us firstly show that $\Delta^\dagger$ is strictly 
positive definite, and therefore admits 
no zero modes by examining the norm squared of a putative zero mode 
\be
|\Delta^\dagger\left(\begin{array}{c}X \\ Y \end{array}\right)|^2
%=\left(\begin{array}{cc} i{\cal D}_z^a & q^{r\dagger} \\ q^r & i{\cal D}_z^f 
%\end{array}\right)\left(\begin{array}{c}X \\ Y \end{array}\right)
&=& |i{\cal D}_z^aX+Yq^\dagger|^2+|Xq-i{\cal D}_{\bar{z}}^fY|^2 \nn\\
&=& |{\cal D}_z^aX|^2+|{\cal D}_{\bar{z}}^fY|^2+|Yq^\dagger|^2 +|Xq|^2=0
\nn\ee
where the vanishing of the cross-terms occurs when evaluated on a 
solution to \eqn{bog}. With all terms on the right-hand side positive 
definite, the last two terms ensure that $X=Y=0$. Thus $\Delta^\dagger$ admits 
no zero modes and ${\cal J}$ counts the number of zero modes of $\Delta$. We now 
turn to the task of evaluating ${\cal J}$. For theories in which the fields 
have suitable fall-off at spatial infinity (faster than $1/r$ in our case -- see the 
second reference in \cite{erick}), 
the quantity ${\cal J}(M^2)$ 
is independent of $M^2$ and the index ${\cal J}$ may be computed more simply in the 
opposite limit $M^2\rightarrow \infty$. It is a simple matter to derive an 
explicit expression for the two composite operators, 
\be
\Delta^\dagger\Delta&=&-\partial_z\partial_{\bar{z}}{\bf 1}_2+
\left(\begin{array}{cc} \Gamma_1-\ft12B^a & L_1 \nn\\
L_2 & \Gamma_2+\ft12B^l\end{array}\right) \nn\\
\Delta\Delta^\dagger&=&-\partial_z\partial_{\bar{z}}{\bf 1}_2 + 
\left(\begin{array}{cc}\Gamma_1 & 0 \nn\\ 0 & \Gamma_2 \end{array}\right)
\nn\ee
where the various operators are defined as,
\be
\Gamma_1X&=&i[\partial_{\bar{z}}A_z,X]+i[A_z,\partial_{\bar{z}}X]
+i[A_{\bar{z}},\partial_zX]-[A_{\bar{z}},[A_z,X]]+Xqq^\dagger \nn\\
\Gamma_2 Y &=& iA_{\bar{z}}\partial_zY+i(\partial_{\bar{z}}A_z)Y+
iA_z\partial_{\bar{z}}Y+A_{\bar{z}}A_zY+Yq^\dagger q \nn\\
L_1Y&=&-iY{\cal D}_z^{\bar{f}}q^\dagger \nn\\
L_2X&=&iX{\cal D}_{\bar{z}}^fq
\nn\ee
Expanding ${\cal J}(M^2)$ in terms of 
$(\nabla+M^2)=(-\partial_z\partial_{\bar{z}}+M^2)$, 
we have
\be
{\cal J}(M^2)=-M^2{\rm Tr}\left[\frac{1}{\nabla+M^2}\left(\begin{array}{cc}
-\ft12 B^a & L_1 \nn\\ L_2 & \ft12 B^l \end{array}\right)\frac{1}{\nabla+M^2}
+\ldots\right]
\nn\ee
where the $\ldots$ vanish in the $M^2\rightarrow\infty$ limit. Taking the trace 
over the adjoint action of $B^a$ causes this term to vanish, and we are left only 
with the left action of $B$ on the space of $N\times N$ matrices $q$. We thus have,
\be
{\cal J}&=&\lim_{M^2\rightarrow\infty}
\sum_{i=1}^N{\rm Tr}\int d^2x\int\frac{d^2k}{(2\pi)^2} 
\frac{-M^2}{(\ft14k^2+M^2)^2}\ft12 B\nn\\
&=& -\sum_{i=1}^N{\rm Tr}\int \frac{d^2x}{2\pi} B \nn\\
&=& Nk
\nn\ee
which counts the complex dimension of $\vkn$ to give the promised result. 

\subsubsection*{The Structure of the Vortex Moduli Space}

Let us now discuss a few basic facts about the vortex moduli space. On 
general grounds, the space decomposes as,
\be
\vkn={\bf C}\times\tvkn
\nn\ee
where ${\bf C}$ parameterises the center of mass of the vortex 
configuration, while information about the relative and internal vortex motion is 
contained within the $2(kN-1)$-dimensional centered vortex moduli space 
$\tvkn$. Supersymmetry, and the BPS-nature of the vortices, ensures that 
the moduli space admits a natural K\"ahler metric defined by the 
overlap of the zero modes,
\be
{\cal L}={\rm Tr} \int d^2x\ \frac{2}{e^2}\dot{A}_z\dot{A}_{\bar{z}}
+\dot{q}\dot{q}^\dagger = g_{a\bar{b}}\dot{z}^a\dot{\bar{z}}{}^{\bar{b}}
\label{manmetric}\ee 
where $z^a$ are complex coordinates on $\vkn$. This is the Manton 
metric, descending from the kinetic terms of the Lagrangian \eqn{lag} 
and is such that geodesics of $g$ describe the classical scattering of vortices 
\cite{manton}.
\para
For the case of the abelian-Higgs model, $N=1$, many 
properties of the vortices and the metric have been studied.
Taubes showed long ago that, as expected, the collective coordinates 
of $\vkn$ correspond to the positions of $k$ unit charge vortices 
moving on the plane and may be identified with the zeros of the Higgs 
field $q$ \cite{taubes}. The metric on ${\cal V}_{k,1}$ can be 
shown  to be geodesically complete and although the exact form of the metric 
remains unknown for $k\geq 2$, several interesting properties were uncovered 
by Samols \cite{samols}. Asymptotically, the metric approaches the 
flat metric on ${\bf C}^k/S_k$ where $S_k$ is the permutation group of 
$k$ elements, reflecting the fact that the vortices are indistinguishable 
particles. The interactions of the vortices 
resolve the orbifold singularities of ${\bf C}^k/S_k$  as the cores overlap.
The leading order corrections to the flat metric,   
which are exponentially suppressed in the separation 
between vortices, were recently calculated by Manton and Speight 
\cite{manspe}. 
\para
The moduli space of vortices in the non-abelian Yang-Mills-Higgs model does 
not appear to have been studied  in the literature. Here we make a few 
elementary remarks. The dimension ${\rm dim}(\vkn)=2kN$ suggests that the 
charge $k$ vortex again decomposes into $k$ unit charge vortices, each of 
which is 
alloted a position on the plane together with $(N-1)$ complex 
internal degrees of freedom describing the orientation of the vortex in 
the $SU(N)_{\rm diag}$ group. Indeed the 
action $SU(N)_{\rm diag}$ on the fields descends to a natural action 
on $\tvkn$, resulting in a holomorphic $SU(N)$ isometry of the metric $g$. 
For $k\geq 2$, there is a further isometry of $\tvkn$ resulting from 
spatial rotations of the vortices. 
\para
Let us examine the moduli space of a single vortex in further detail. 
Given a specific solution $(B_\star, q_\star)$ to the abelian vortex equations, 
one can always construct a solution to the non-abelian equations \eqn{bog} by 
simply embedding $(B_\star, q_\star)$ in the upper-left corner 
of the $N\times N$ matrices $B$ and $q$. In the case of a single vortex 
$k=1$, acting on this configuration with the $SU(N)_{\rm diag}$ symmetry 
sweeps out the full moduli space of solutions. Since the vortex embedded 
in the upper-left corner breaks $SU(N)_{\rm diag}\rightarrow SU(N-1)\times U(1)$, 
the vortex moduli space is
\be
\tilde{\cal V}_{1,N}\cong SU(N)/(SU(N-1)\times U(1))\cong {\bf CP}^{N-1}
\nn\ee
endowed with the round Fubini-Study metric. The only information that 
we still need to determine is the overall scale of the moduli space. 
This will be important later in matching to the instanton moduli space. 
Since ${\bf CP}^{N-1}$ is a homogeneous space, we can fix the scale by 
calculating the overlap of {\it any} two suitable zero modes arising from 
the $SU(N)_{\rm diag}$ action. For $\Omega(z,\bar{z})\in su(N)$, the zero modes 
associated to an $SU(N)_{\rm diag}$ rotation are given by,
\be
\dot{A}={\cal D}^a\Omega\ \ \ ,\ \ \ \dot{q}=i(\Omega q-q\Omega_0)
\label{zeromodes}\ee
where $\Omega\rightarrow\Omega_0$ as $|z|\rightarrow\infty$. The 
transformation of $q$ arises because the left action is by the 
$U(N)_G$ gauge symmetry, while the right action is by the $SU(N)_F$ 
flavour symmetry.  The $z$ dependence of $\Omega$ is required in order to 
satisfy the gauge fixing condition \eqn{gauge} which becomes
\be
({\cal D}^a)^2\Omega=e^2(\{\Omega,qq^\dagger\}-2q\,\Omega_0\,q^\dagger)
\nn\ee
For the initial configuration embedded in the upper-left corner of $B$ and 
$q$, these equations are solved by the $(N-1)$ rotations, 
\be
(\Omega^a_{\ b})^j=\left(\frac{q_\star}{\sqrt{\zeta}}\right)
\delta^a_{\ 1}\delta_{\ b}^j
+ \left(\frac{q_\star^{\dagger}}{\sqrt{\zeta}}\right)
\delta^{ja}\delta_{b1}
\ \ \ \ \ j=2,\ldots,N
\nn\ee
and it is a simple matter to compute the overlap \eqn{manmetric} of the 
zero modes \eqn{zeromodes} to determine the overall radius of the moduli 
space to be
\be
{\rm Radius}^2\left(\tilde{\cal V}_{1,N}\right)\sim\frac{1}{e^2}
\label{vortsize}\ee
Finally, let us make a brief comment on the spectrum of vortices in the quantum 
theory. In the $d=2+1$ theory with ${\cal N}=4$ supersymmetry, ground states of 
the vortices in a given sector are associated to harmonic forms on $\tvkn$. For 
the case of a single vortex, there are therefore $\chi({\bf CP}^{N-1})=N$ 
such states, implying that the vortex transforms in the fundamental 
representation of $SU(N)_{\rm diag}$.

\section{Branes}

In this section, we discuss a brane realisation of the vortices in type IIB 
string theory. We start with 
the $d=2+1$, ${\cal N}=4$ $U(N)$ Yang-Mills-Higgs theory described in 
the Lagrangian \eqn{lag}. The brane realisation of this is well known \cite{hw} and consists 
of $N$ D3 branes,  suspended between two parallel NS5-branes. A further $N$ 
semi-infinite $D3$ branes connect to the right-hand NS5-brane to provide the 
hypermultiplets. 
%%%%%%%%%%%%%%%%%%%%%%%%%%%%%%%%%%%%%
\newcommand{\onefigure}[2]{\begin{figure}[htbp]

         \caption{\small #2\label{#1}(#1)}
         \end{figure}}
\newcommand{\onefigurenocap}[1]{\begin{figure}[h]
         \begin{center}\leavevmode\epsfbox{#1.eps}\end{center}
         \end{figure}}
\renewcommand{\onefigure}[2]{\begin{figure}[htbp]
         \begin{center}\leavevmode\epsfbox{#1.eps}\end{center}
         \caption{\small #2\label{#1}}
         \end{figure}}
%%%%%%%%%%%%%%%%%%%%%%%%%%%%%%%%%%%%
\begin{figure}[htb]
\begin{center}
\epsfxsize=6in\leavevmode\epsfbox{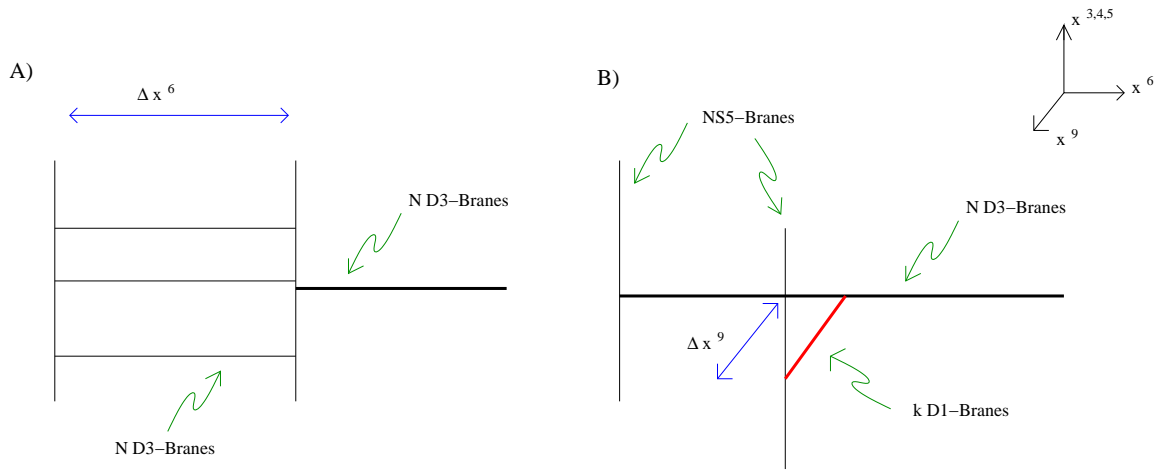}
\end{center}
\caption{The brane configuration for $U(N)$ gauge theory with $N$ 
hypermultiplets. Figure 1A shows the theory on the 
Coulomb branch. In Figure 1B, the theory has a FI parameter and lies 
in its unique ground state. The D1-branes are the vortices.}
\label{vibfig1}
\end{figure}
\para
In Figure 1 we draw this brane configuration, firstly on the Coulomb branch 
with $\zeta=0$, and secondly on the Higgs branch in which one NS5-brane is 
separated from the other branes, inducing a non-zero FI parameter $\zeta$. 
In the second picture, we also include the BPS vortices which appear as 
$k$ D1-branes stretched between the D3-branes and the isolated NS5-brane. To 
see that these D1-branes are indeed identified with vortices, note that they 
are the only BPS states of the brane configuration with the correct mass. 
The spatial worldvolume directions of the branes follow official convention:
\be
NS5: && 12345 \nn\\
D3: && 126 \nn\\
D1: && 9
\nn\ee
Both the FI parameter $\zeta$, and the gauge coupling, are encoded in the 
separation $\Delta x$ between the two NS5-branes. We have
\be
\frac{1}{e^2}=\frac{\Delta x^6}{2\pi g_s}\ \ \ \ \ ;\ \ \ \ \ \ \ 
\zeta=\frac{\Delta x^9}{4\pi^2 g_s l^2_s}
\label{rel1}\ee
where $l_s=\sqrt{\alpha^\prime}$ 
and $g_s$ are the string length and coupling respectively. 
To take the gauge theory decoupling limit, we want to 
send $g_s\rightarrow 0$, while insisting that the field 
theory excitations are much smaller than other stringy and 
Kaluza-Klein modes. The two mass scales of the field theory are 
the mass of the photon $M_\gamma\sim\sqrt{e^2\zeta}$ and the 
mass of the vortex $M_v\sim \zeta$. An interesting curiosity about 
vortices is that while their mass is $M_v$, their size is $M^{-1}_\gamma$. 
In order to decouple the gauge theory from the string dynamics, we 
require
\be
M_\gamma\, ,\,M_v \ll 1/l_s\, ,\, 1/\Delta x^6
\nn\ee
while the ratio 
$(M_v/M_\gamma)^2\sim\Delta x^6 \Delta x^9 /l_s^2g_s^2$ remains fixed. The 
decoupling limit can therefore 
be achieved by setting $\Delta x^6\sim \epsilon l_s$ and 
$\Delta x^9\sim \epsilon^3l_s$ and $g_s\sim\epsilon^2$, taking $\epsilon\rightarrow 0$.
\para
Let us now turn to the vortices. It is a simple matter 
to read off the theory living on the worldvolume of the D1-branes 
(similar configurations were 
considered previously in the T-dual picture \cite{bh,hh}). The dynamics of the 
D1-branes is controlled by an ${\cal N}=(2,2)$ supersymmetric, gauged quantum 
mechanics. The relevant representations of the supersymmetry algebra are 
simply the dimensional reduction of the familiar vector and chiral multiplets 
in $d=3+1$ dimensions. The vortex theory involves 
a $U(k)$ vector multiplet, consisting of a gauge field together with three 
adjoint scalar fields $\phi^r$, $r=1,2,3$ parameterising the motion of 
the D1-branes in the $x^{r+2}$ directions. 
These are coupled to an adjoint chiral multiplet whose complex scalar we 
denote $Z$. The eigenvalues of $Z$ parameterise the position of the 
$k$ D1-branes in the $z=x^1+ix^2$ plane. A further $N$ fundamental 
chiral multiplets, with complex scalars $\psi$, arise from the D1-D3 
strings. The global symmetry group of the theory is
\be
G=SU(2)_R\times SU(N)_D\times U(1)_F
\label{vortsym}\ee
where $SU(2)_R$ is an R-symmetry rotating the scalars in the vector 
multiplet\footnote{For vortex solutions whose worldvolume is 
$d$-dimensional, this R-symmetry group is $Spin(4-d)$.}, $U(1)_F$ 
is a flavour symmetry rotating the phase of $Z$ and $SU(N)_D$ is a 
flavour symmetry acting on $\psi$ in the anti-fundamental 
representation. The $\psi$ fields may be represented as 
$k\times N$ matrices, with 
the $U(k)$ gauge group acting by left multiplication, and the 
$SU(N)_D$ flavour symmetry acting by right multiplication. We use the notation
\be
\psi=\psi^m_{\ i}\ \ \ \ \ \ \ \ \ \ \ m=1,\ldots,k\ ;\ i=1,\ldots, N
\nn\ee
All of these fields come with fermionic superpartners which we suppress. 
The bosonic Lagrangian is given by
\be
{\cal L}_{\rm vort}&=&{\rm Tr}\left[\frac{1}{2g^2}{\cal D}_t\phi^r{\cal D}_t\phi^r
+{\cal D}_tZ^\dagger{\cal D}_tZ + 
{\cal D}_t\psi_i {\cal D}_t\psi^{i\dagger} -\frac{1}{2g^2}[\phi^r,\phi^s]^2
\right.\nn\\
&&\left.\ \ \ \ \ -|[Z,\phi^r]|^2-\psi\psi^\dagger\phi^r\phi^r
-\frac{g^2}{2}\left(\psi\psi^\dagger-[Z,Z^\dagger]-r{\bf 1}_k\right)^2
\right]
\label{vtheory}\ee 
Once again, the gauge coupling $g^2$ and FI parameter $r$ of this theory 
are determined by the separation of the NS5-branes, although with 
reciprocal relations to the D3-brane theory \eqn{rel1}
\be
\frac{1}{g^2}=\frac{2\pi l_s^2\,\Delta x^9}{g_s}\ \ \ \ \ ;\ \ \ \ \ \ \ 
r=\frac{\Delta x^6}{g_s}
\label{rel2}\ee
We see that taking the decoupling limit of the D3-brane theory implies 
the strong coupling limit of the vortex theory $g^2\rightarrow\infty$. 
However, the FI parameter $r$ remains finite and in fact is identified 
with the gauge coupling $e^2$
\be
r=\frac{2\pi}{e^2}
\label{re2}\ee
For $r\neq 0$, there is no Coulomb branch, so that taking 
the strong coupling limit $g^2\rightarrow\infty$ decouples the 
vector multiplet fields $\phi^r$ and restricts attention to the 
Higgs branch of the theory. We shall denote this Higgs branch as 
${\cal M}_{k,N}$. It is given by a $U(k)$ K\"ahler quotient of 
${\bf C}^{k(N+k)}$, parameterised by $Z$ and $\psi$. The 
associated moment map is simply the D-term from \eqn{vtheory}
\be
D^m_{\ n}=\psi^m_{\ i}\psi^{i\dagger}_{\ n}-[Z,Z^\dagger]^m_{\ n}-
r\delta^m_{\ n}=0
\label{D}\ee
This imposes $k^2$ real constraints on ${\bf C}^{k(N+k)}$, while modding 
out by the $U(k)$ gauge group reduces the dimension of the Higgs branch 
by another $k^2$. Thus the real dimension of the Higgs branch is
\be
{\rm dim}({\cal M}_{k,N})=2kN
\nn\ee
which we recognise as the dimension of the vortex moduli space \eqn{dimvkn}. 
Indeed, the main result of this paper is the brane-predicted isomorphism
\be
\vkn\ \cong\ {\cal M}_{k,N}
\label{bigwin}\ee

\subsubsection*{Some Examples and the Metric}

Let us examine the claim \eqn{bigwin} in more detail. Firstly, note that the center of 
mass position of the D1-branes, given by $Z=z{\bf 1}$, decouples from 
the other fields, guaranteeing that the Higgs branch decomposes as 
\be
\mkn\ \cong\ {\bf C}\times \tilde{\cal M}_{k,N}
\nn\ee
in agreement with the vortex moduli space. To make further comparisons, 
let us consider specific examples, starting with the description of 
a single vortex $k=1$ in the $U(N)$ theory. In this case the vortex dynamics 
is abelian so $Z$ decouples 
and the D-term constraint reduces to $|q|^2=r$ where $q$ is an 
$N$-vector. We are left with the well known gauged linear sigma-model 
construction of ${\bf CP}^{N-1}$, and we have,
\be
\tilde{{\cal M}}_{1,N}\ \cong\ {\bf CP}^{N-1}\ \cong\  \tilde{{\cal V}}_{1,N}
\nn\ee
The size, or K\"ahler class, of the Higgs branch is determined by the FI parameter 
$r=2\pi/e^2$ in agreement with the vortex moduli space \eqn{vortsize}.
\para
The second example we consider is that of $k$ vortices in the abelian-Higgs 
model with $N=1$. This vortex quantum mechanics was previously studied in 
\cite{lee} as a matrix model for identical particles moving on the plane. 
Prior to that, the D-term constraints \eqn{D} were 
solved in a somewhat different context by Polychronakos \cite{poly}, who 
showed that a given solution is uniquely determined by a set of 
eigenvalues for $Z$, up to Weyl permutations. Thus
\be
{\cal M}_{k,1}\ \cong\ {\bf C}^k/S_k\ \cong\ {\cal V}_{k,1}
\nn\ee
In these two, simple cases, we have therefore confirmed that the Higgs branch 
and vortex moduli spaces are indeed isomorphic. We now turn to the question 
of the metric. The Higgs branch  
$\mkn$ inherits a natural K\"ahler metric from the K\"ahler 
quotient construction described above. The presence of the flavour symmetry $SU(N)_D$ 
guarantees that this metric exhibits an $SU(N)$ holomorphic isometry. 
For $k\geq 2$, $\tilde{\cal M}_{k,N}$ also enjoys a $U(1)$ holomorphic 
isometry, arising from $U(1)_F$, corresponding 
to rotating the branes in the $x^1+ix^2$ plane. Thus the quotient metric 
on $\mkn$ and the Manton metric on $\vkn$ share the same isometries. 
Indeed, from the brane picture it is clear that the $SU(N)_{\rm diag}$ 
and $SU(N)_D$ symmetry groups of the D3-brane and D1-brane theories, 
share the same origin.
\para
Do further properties of the metrics coincide? In the case of $k=1$, the 
metric on $\tilde{\cal M}_{1,N}$ is the round Fubini-Study metric on 
${\bf CP}^{N-1}$, in agreement with the Manton 
metric on $\tilde{\cal V}_{1,N}$. However, in this case the agreement 
is a consequence of the symmetries of the problem. In general, the 
metrics are not the same. To see this, let us return to the case of the 
abelian-Higgs model with $N=1$. Importantly, the asymptotic metric on 
${\cal M}_{k,1}$ is the flat metric on ${\bf C}^k/S_k$, in agreement 
with the Manton metric. This is crucial 
to ensure that the Higgs branch describes the moduli space of indistinguishable 
particles since mere topological information does not suffice (topologically 
${\bf C}^k/S_k\cong{\bf C}^k$ as any polynomial will confirm). However, in 
the case of the K\"ahler quotient, the leading order corrections to the 
flat metric are power-law. This is to be contrasted with the exponential 
corrections of the Manton metric. To be concrete, consider the case 
$k=2$, $N=1$. The metrics on both $\tilde{\cal V}_{2,1}$ and 
$\tilde{\cal M}_{2,1}$ take the form,
\be
ds^2=f^2(\sigma)(d\sigma^2+\sigma^2d\theta^2)
\label{cone}\ee
where $\sigma$ is the separation between vortices, or D1-branes, and 
$\theta\in[0,\pi)$ so that the moduli space looks like a cone. For the Higgs 
branch, the explicit K\"ahler quotient construction was performed in 
\cite{lee} and the conformal factor is given by,
\be
f_{\cal M}^2(\sigma)=\frac{\sigma^2}{\sqrt{\sigma^4+r^2}}\approx 1 - 
\frac{r^2}{2\sigma^4}+\ldots
\label{2vort}\ee
The calculation of the leading order scattering of vortices was performed 
in \cite{manspe}, and the equivalent metric on $\tilde{\cal V}_{2,1}$ 
was computed to be,
\be
f_{\cal V}^2(\sigma)\approx 1-\lambda^2\sqrt{\frac{4\pi}{\sigma}}e^{-2\sigma}
+\ldots
\label{man2vort}\ee
where $\lambda$ is a coefficient which parameterises the asymptotic 
return to vacuum of the Higgs field in the solution to \eqn{bog}. 
This coefficient is not known analytically but it was shown in 
\cite{tong} that T-duality between the $A_N$ singularity and 
{\em fully localised} NS5-branes requires a worldsheet instanton effect  
and holds only if $\lambda=8^{1/4}\approx 1.682$. 
This is in agreement with the numerical result $\lambda \approx 1.683$ 
of \cite{spe}. To summarise, we see that, while the metrics on ${\cal V}_{k,1}$ and 
${\cal M}_{k,1}$ are asymptotically, and 
qualitatively, similar they differ in the details.

\section{Instantons}

The K\"ahler quotient construction of the vortex moduli space is reminiscent 
of the hyperK\"ahler quotient of the moduli space of $k$ instantons in 
$U(N)$ Yang-Mills theory. We denote this latter space as $\ikn$. In this 
Section, we make the connection between $\vkn$ and $\ikn$ more explicit. 
We begin with a review of the ADHM gauge theory describing instantons 
on non-commutative ${\bf R}^4$, with the specific 
anti-self-dual, commutation relations
\be
[x^1,x^2]=i\theta\ \ ,\ \ [x^3,x^4]=-i\theta
\label{noncom}\ee
with all other commutators vanishing. 
Recall that the ADHM construction of $\ikn$, as proposed in \cite{adhm},  
can be elegantly described in terms of an auxiliary $U(k)$ gauge 
theory with 8 supercharges \cite{adhmgauge}. The matter content of 
this theory includes an adjoint valued hypermultiplet and $N$ fundamental 
hypermultiplets. The instanton moduli space is described as a hyperK\"ahler 
quotient as the Higgs branch of this gauge theory, parameterised by the 
hypermultiplet scalar fields. Denote the two complex scalars in the adjoint 
multiplet as $Z$ and $W$, and the $2N$ complex scalars in the remaining 
hypermultiplets as $\psi$ and $\tilde{\psi}$. While $\psi$ transforms in the 
${\bf k}$ representation of the gauge group, $\tilde{\psi}$ transforms as 
$\bar{\bf k}$, and we represent both of these fields 
as a $k\times N$ (respectively $N\times k$) matrix,
\be
\psi=\psi^m_{\ i}\ \ ,\ \ \tilde{\psi}=\tilde{\psi}^i_{\ m},\ \ \ \ \ \ \ 
\ \ \ \ \ \ \ \ m=1,\ldots,k,\ i=1,\ldots, N
\nn\ee
Theories with 8 supercharges have a triplet of D-terms which, in 4 supercharge 
language, can be decomposed into a D-term and F-term. These constraints, which 
provide the triplet of moment maps in the hyperK\"ahler quotient construction, read
\be
\begin{array}{c}
D^m_{\ n}=\psi^m_{\ i}\psi^{i\dagger}_{\ n}-\tilde{\psi}^{m\dagger}_{\ i}
\tilde{\psi}^i_{\ n}-[Z,Z^\dagger]^m_{\ n}-[W,W^\dagger]^m_{\ n}-r\delta^m_{\ n}=0 
\nn\\
F^m_{\ n}=\psi^m_{\ i}\tilde{\psi}^i_{\ n}+i[Z,W]^m_{\ n}=0
\end{array}
\nn\ee
The FI parameter $r$ appears only in the D-term, a fact related to the specific 
choice of non-commutative background \eqn{noncom} as shown by Nekrasov and 
Schwarz \cite{ns}. The relationship is simply
\be
r=4\theta
\nn\ee
The role of $r$ is to resolve the singularities of $\ikn$ 
in the manner proscribed by Nakajima \cite{nak}. In doing so, it picks out a 
preferred complex structure on $\ikn$. 
\para
Note that we have used the same 
notation in the ADHM gauge theory as we did in the 
the previous section, and we will shortly explain the deformation which 
takes us from ADHM to the vortex theory. Before doing so, it will do us well 
to dwell a little on the symmetries of the ADHM theory. To compare with 
the previous section we choose to define the ADHM theory in $d=0+1$ 
dimensions, describing particles in $d=4+1$ dimensional Yang-Mills or, 
alternatively, D0-branes moving in the background of D4-branes. 
The global symmetry group of the ADHM theory is 
\be
G^\prime=Spin(5)_R\times SU(N)_F \times SU(2)_L \times U(1)_R
\label{gprime}\ee
The $Spin(5)_R$ symmetry rotates the scalars in the vector multiplet\footnote{For 
instantons with a $d$-dimensional worldvolume, this R-symmetry group is 
$Spin(6-d)$.}. The $U(1)_R\times SU(2)_L$ is what remains of the $SO(4)$ 
spatial rotation group of ${\bf R}^4$ with the anti-self-dual 
non-commutative deformation \eqn{noncom}, and the $SU(N)_F$ descends from 
the $U(N)$ gauge symmetry on the D4-branes.  
The adjoint doublet $(Z,W)$ 
transforms as $({\bf 1},{\bf 1},{\bf 2})_{+1}$ under $G^\prime$, 
while $\psi$ transforms 
as $({\bf 1},\bar{\bf N},{\bf 1})_{+1}$ and $\tilde{\psi}$ transforms 
as $({\bf 1},{\bf N},{\bf 1})_{+1}$, where the subscripts denote the 
charge $Q_R$ under the $U(1)_R$ R-symmetry. 
\para
We are now in a position to describe the deformation which takes us 
to the vortex theory by adding masses to all the unnecessary fields. 
We accomplish this 
by {\it weakly gauging} a particular $\widehat{U(1)}$ symmetry. This involves 
gauging a symmetry in a manner consistent with supersymmetry. The scalars 
in this new vector multiplet are then endowed with vacuum expectations 
values (vevs) and the new vector multiplet is subsequently decoupled. The 
only remnant of the whole process is the vevs, which give masses to any 
field charged under the $\widehat{U(1)}$ symmetry. If $\widehat{U(1)}$ is 
taken to be a flavour symmetry, then this process preserves the full 8 supercharges 
of the ADHM theory. In contrast, if $\widehat{U(1)}$ is a generic 
R-symmetry, this process breaks all supersymmetry. However, there are 
specific combinations of R-symmetries which one may gauge which preserve  
a fraction of the supersymmetry and it is this combination that we shall 
employ. Let $U(1)_A\subset Spin(5)$ be such that it rotates two of the 
vector multiplet scalars, leaving the remaining three untouched;  
let $U(1)_L\subset SU(2)_L$ have the Pauli matrix generator $\tau^3$; and 
let $U(1)_G\subset U(k)$ be the overall gauge rotation. 
Then we choose the combination of symmetries that act on fields with 
charge $Q$, such that
\be
\widehat{Q}=Q_A+Q_R-Q_L-Q_G
\label{qhat}\ee
The fields $W$ and $\tilde{\psi}$, together with two of the five vector 
multiplet scalars, have $\widehat{Q}\neq 0$. These all receive 
masses. The fields $Z$ and $\psi$, and the three remaining scalars of the 
vector multiplet all have $\widehat{Q}=0$ and survive unscathed. 
We are left with the vortex theory of Section 2, with the relationship 
between the FI parameter and parameters giving,
\be
\theta=\frac{\pi}{2e^2}.
\nn\ee

\subsubsection*{The Moduli Spaces}

While the above discussion has been in terms of the ADHM gauge theory, 
the deformation also has a simple description directly in terms 
of the instanton moduli space $\ikn$. 
The $\widehat{U(1)}$ symmetry of the gauge theory descends to an 
$\hat{\bf S}^1$ 
action on $\ikn$, endowing the metric on $\ikn$ with a Killing 
vector $\hat{k}$. This Killing vector is holomorphic, preserving 
the preferred complex structure while rotating the remaining two. 
The mass terms introduced above by {\it weakly gauging} 
$\widehat{U(1)}$ induce to a potential $V$ on $\ikn$ proportional to 
the length${}^2$ of the Killing vector,
\be
V\sim \hat{k}^2
\nn\ee
Such potentials have been widely used in soliton physics recently 
(see for example \cite{pot},\cite{potinst}), although usually in the context 
of supersymmetry-preserving tri-holomorphic Killing vectors. We 
therefore have a description of the vortex moduli space 
$\vkn$ directly in terms of the instanton moduli space
\be
\left.\begin{array}{c}\! \end{array}{\cal V}_{k,N}\ \cong\ 
{\ikn}\ \right|_{\hat{k}=0}
\nn\ee
The zeroes of the Killing vector $\hat{k}$ are precisely the fixed points 
of the $\hat{\bf S}^1$ action. The meaning of this action can be 
determined from the assignment of charges $\hat{Q}$ in \eqn{qhat}. 
Recall that $U(1)_R\times U(1)_L\subset SU(2)_R\times U(1)_L$ is 
the subgroup of the rotations $SO(4)\cong SU(2)_R\times SU(2)_L$ of 
${\bf R}^4$ that are left unbroken by the non-commutative deformation 
\eqn{noncom}. We find therefore that the action $\hat{Q}$
corresponds to rotating the instantons in the $x^3-x^4$ plane, and the 
vortices are related to instantons which are invariant under this 
$U(1)$ action.
\para
Let us now turn to some examples: the moduli space $\tilde{\cal I}_{1,N}$ 
of a single instanton in $U(N)$ non-commutative Yang-Mills is given 
by the cotangent bundle $T^\star ({\bf CP}^{N-1})$ 
endowed with the Calabi metric \cite{leeyi}. The potential $\hat{k}^2$ 
vanishes on the zero section of the bundle ${\bf CP}^{N-1}$, reducing 
to the moduli space of a single vortex $\tilde{\cal V}_{1,N}$. 
Another example: the moduli space $\tilde{\cal I}_{2,1}$ of 
two instantons in $U(1)$ gauge theory is the Eguchi-Hanson metric on 
$T^\star ({\bf S}^2)$. The explicit hyperK\"ahler quotient construction 
was performed in \cite{ksme}. Note that this case is special 
since $\tilde{\cal I}_{2,1}\cong\tilde{\cal I}_{1,2}$, which is 
not true for $k>2$. However, the tri-holomorphic $SU(2)$ isometry 
of $T^\star({\bf S}^2)$ has a different origin in these two cases. In the 
notation of \eqn{gprime}, the isometry is $SU(2)_F$ for $\tilde{\cal I}_{1,2}$, 
while it is $SU(2)_L$ for $\tilde{\cal I}_{2,1}$. Since, from \eqn{qhat}, 
the potential on the instanton moduli space involves $SU(2)_L$, but not $SU(2)_F$, 
the vortex moduli spaces $\tilde{\cal V}_{2,1}$ and $\tilde{\cal V}_{1,2}$ 
are given by different holomorphic submanifolds of $T^\star({\bf S}^2)$. 
It is a simple exercise to show that the vacua of the potential on 
$\tilde{\cal I}_{2,1}$ is the two dimensional cone endowed with the 
metric \eqn{2vort}. 

\subsubsection*{A Wrapped Brane Realisation}

From the perspective of the D4-brane, the above deformation of the 
instanton theory involves locking the $U(1)_{L/R}$ symmetries tangent 
to the D4-brane, with the $U(1)_A$ symmetry normal to the D4-branes. 
This is reminiscent of the twisting of the tangent and normal bundles 
of branes when wrapped on cycles \cite{vetal}. In this 
section, we give evidence suggesting that the two are indeed related.
\para
To see this connection, let us first return to the brane set-up of 
Section 3 as depicted in Figure 1. We perform a T-duality in 
the $x^9$ direction, and describe the resulting IIA string theory 
set-up. Under T-duality, the two NS5-branes are replaced 
by the background geometry ${\bf C}^2/{\bf Z}_2$. (The duality between NS5-branes 
and $ALF$ spaces was 
first conjectured by Hull and Townsend \cite{ht}. A proof from the 
worldsheet sigma-model, including the breaking of translation symmetry 
associated to the localization of the NS5-brane, was given in \cite{tong}).
The separation of the NS5-branes in the $x^6$ direction resolves 
the orbifold singularity, resulting in the background spacetime  
$T^\star({\bf S}^2)$\footnote{Note that this ubiquitous space has already 
appeared twice as the instanton moduli spaces $\tilde{\cal I}_{1,2}$ 
and $\tilde{\cal I}_{2,1}$. Here it appears in an 
unrelated context as the background spacetime in string theory.}. 
Topologically, this space can be thought of as an ${\bf S}^1$ 
fibration, parameterised by $x^9$, over ${\bf R}^3$, parameterised by 
${\bf r}=(x^6,x^7,x^8)$. In Gibbons-Hawking coordinates, the metric 
takes the form,
\be
ds^2 = H(r)\,dr^2 + \ft14 H(r)^{-1}((dx^9)^2 + \bomega\cdot d{\bf r})^2
\nn\ee
where $\nabla\times\bomega = \nabla H$ and 
\be
H(r)=\frac{1}{|{\bf r}-{\bf r}_0|}+\frac{1}{|{\bf r}+{\bf r}_0|}
\nn\ee
The 3-vector ${\bf r}_0$ resolves the orbifold singularity and, for 
the T-dual of Figure 1, is given by ${\bf r}_0\sim(1/e^2,0,0)$. 
The ${\bf S}^1$ fiber degenerates at the two points 
${\bf r}=\pm{\bf r}_0$, resulting in  the Christmas cracker topology 
shown in Figure 2. The zero section ${\bf S}^2$, which can be clearly 
seen in this picture, contains a paper hat and a 20 year old joke. 
\begin{figure}[htb]
\begin{center}
\epsfxsize=4in\leavevmode\epsfbox{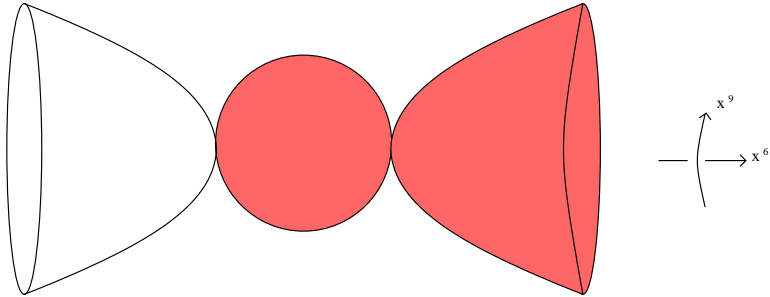}
\end{center}
\caption{The Eguchi-Hanson Christmas cracker. The D4-branes are 
wrapped around the shaded region.}
\label{vibfig2}
\end{figure}

What becomes of the D-branes after T-duality?
The D3-branes of Figure 1 become D4-branes with 
worldvolume spanning $x^1,x^2,x^6$ and $x^9$. 
They wrap the 
compact $S^2$, and one half of the cracker as depicted by shading 
in the Figure. The vortices 
are a little more mysterious. Had the D1-branes in Figure 1 been 
infinite in the $x^9$ direction, they would become D0-branes in 
the IIA description. Since the D1-branes actually stretch only a fraction 
of the distance, we expect that they become fractional D0-branes. 
However, such objects are usually understood in terms of a 
D2-$\bar{\rm D2}$ pair wrapping a vanishing ${\bf S}^2$, 
through which an NS-NS B-field threads in order to provide the 
D0-brane charge. Yet in our case the ${\bf S}^2$ has finite size, 
and such an interpretation breaks down, as can easily be seen by 
computing the mass of such a D2-$\bar{\rm D2}$ pair. It would be 
interesting to get a better understanding of these fractional 
D0-branes in this picture, and complete the relationship to the wrapped 
D0-D4 system.

%However, there remains a puzzle. The radius of the spatial 
%${\bf S}^2$ is proportional to $\Delta x^6 \sim 1/e^2$. In 
%IIB string theory, the NS5-branes are also separated in 
%the $x^9$ direction, which becomes a NS-NS B-field under T-duality 
%\cite{who}. More specifically, $T^\star({\bf S}^2)$ contains an 
%anti-self-dual harmonic 2-form $\omega_2$, and we have 
%\be
%B_2\sim (\Delta x^9)\omega_2
%\nn\ee
%It is expected that this B-field results in a non-commutative 
%deformation $\theta$ of the worldvolume theory on the D4-brane 
%\cite{sw}.  
%However, from the previous section we have seen that 
%$\theta\sim 1/e^2\sim \Delta x^6$. It remains to be seen how 
%this detail works out.

\section{Semi-Local Vortices and Sigma-Model Lumps}

In this Section, we discuss a generalisation of the vortices 
to $U(N)$ Yang-Mills with $N_f=(N+M)$ flavours. The 
Lagrangian takes the same form as previously \eqn{lag} except the 
fundamental scalars are now $N\times (N+M)$ dimensional matrices
\be
q=q^a_{\ i}\ \ ,\ \ \tilde{q}=\tilde{q}^i_{\ a}\ \ \ \ \ \ \ \ a=1\ldots, N\ ,\ 
i=1,\ldots, N+M
\nn\ee
Rather than the unique, isolated vacuum of Section 2, the theory now 
has a Higgs branch of vacua. However, if $\tilde{q}$ develops 
an expectation value, then there are no BPS vortex solutions. This 
may be easily seen from an analysis of the Bogomoln'yi equations, 
and follows from the mathematical fact truth that a line bundle of 
negative degree admits no holomorphic sections (see, for example, 
\cite{wittphase} for the translation). We therefore restrict 
attention to the reduced Higgs branch of vacua, denoted ${\cal N}_{N,M}$ 
obtained by insisting $\tilde{q}=0$. For example, for abelian 
theories with $N=1$, the Higgs branch of vacua is the cotangent bundle 
$T^\star({\bf CP}^M)$, while the reduced Higgs branch describing the 
vacua which admit BPS vortex solutions 
is simply ${\cal N}_{1,M}={\bf CP}^M$. 
In general, the reduced Higgs branch is the Grassmannian of $N$ 
planes in ${\bf C}^{N+M}$,
\be
{\cal N}_{N,M}=G(N,N+M)
\nn\ee
This is a symmetric space, and we may choose to work in any of the 
vacua without loss of generality. We pick,
\be
\begin{array}{ll}
q^a_{\ i}=\sqrt{\zeta}\delta^a_{\ i}\ \ \ \ \ \ \ \ & i=1,\ldots N 
\nn\\
q^a_{\ i}=0 & i=N+1,\ldots N+M \nn\\
\tilde{q}^i_{\ a}=0 & i=1,\ldots, N+M
\end{array}
\nn\ee
In this vacuum the $SU(N+M)_F$ flavour symmetry of the theory is 
broken in the pattern,
\be
U(N)_G\times SU(N+M)_F\rightarrow S\left[U(N)_{\rm diag}\times U(M)_F\right]
\label{survivor}\ee
The theory admits BPS vortices with the Bogomoln'yi equations taking 
the same form as previously \eqn{bog} with $q$ now 
interpreted as a matrix of the appropriate size. 
%Writing the indices explicitly, we have
%\be
%B^a_{\ b}&=&e^2(q^a_{\ i}q^{i\dagger}_{\ b}-\zeta\delta^a_{\ b}) \nn\\
%({\cal D}_zq_{i})^a&=&0
%\label{semibog}\ee
We denote the moduli space of vortices in this model as 
$\hat{\cal V}_{k,(N,M)}$. Note that, in the notation of the previous 
sections, we have $\vkn\cong\hat{\cal V}_{k,(N,0)}$. 
It is a simple matter to generalise the index theorem of Section 2 to 
the present case. We omit the details, stating only the result 
${\cal J}(M^2\rightarrow\infty) = k(N+M)$. We therefore have
\be
{\rm dim}(\hat{\cal V}_{k,(N,M)})=2k(N+M)
\label{hatdim}\ee
Note that since we have taken the limit $M^2\rightarrow\infty$, this computation 
ignores a surface term contribution which comes from fields dropping off as 
$1/r$ \cite{erick}. Indeed, as we shall review below, 
it is known that the counting \eqn{hatdim} includes zero modes which are not 
$L^2$ normalisable and which one would not, therefore, expect to be included in 
${\cal J}={\cal J}(M^2=0)$. Nevertheless, these modes corresponds to collective 
coordinates of the semi-local vortex and we wish to keep them in our 
discussion, so the result \eqn{hatdim} is the relevant one.  
\para
In the abelian case $N=1$, the vortex equations with multiple Higgs fields 
have been well studied 
in the literature, where they go by the name of {\em semi-local vortices}. 
For a review of their properties and their relationship to electroweak strings, 
see \cite{vash}. The result ${\rm dim}(\hat{\cal V}_{k,(1,M)})=2k(1+M)$ was 
previously determined from a direct analysis of the Bogomoln'yi equations 
in \cite{gibbons} and subsequently from a brane picture in \cite{tongdw}. 
\para
An interesting feature of the semi-local vortices is 
that they may remain non-singular in the 
limit $e^2\rightarrow\infty$. This is in contrast to the 
vortices considered in Section 2 whose size scales as $(e^2\zeta)^{-1/2}$ 
and thus become point-like objects in this limit.  
In fact, the semi-local vortices reduce to another familiar 
topological object as $e^2\rightarrow\infty$: they become 
sigma-model lumps (a.k.a sigma model instantons, or 
textures) on the reduced target space ${\cal N}_{N,M}$. While the 
vortices are supported by $\Pi_1(U(N))$, the lumps are 
supported by $\Pi_2({\cal N}_{N,M})$.
For example, the semi-local vortices 
of the abelian $N=1$ model become lumps on ${\bf CP}^{M}$. 
A nice description of how the metamorphosis from vortex to lump 
occurs may be found in  \cite{bert}. Thus, in the limit 
$e^2\rightarrow\infty$, the moduli space of semi-local 
vortices $\hat{\cal V}_{k,(N,M)}$ becomes the moduli space 
of Grassmannian $G(N,N+M)$ lumps.
\para
It is well known that sigma-model lumps share several properties with 
Yang-Mills instantons. In particular, they may have arbitrary size 
and, in the $e^2\rightarrow\infty$ limit, $k$ of the 
collective coordinates of \eqn{hatdim} may be thought of as the 
scales of the $k$ lumps. Since the lumps may have any size, they can 
also shrink to a singular solution. Thus, 
just like the unresolved instanton moduli spaces, the moduli space of 
lumps contains singularities. 
These singularities are removed by introducing a 
gauge field with a finite coupling $e^2$ and returning to the full 
vortex equations \eqn{bog}. In this way, the inverse gauge coupling 
$1/e^2$ plays a role in the vortex dynamics reminiscent of the 
non-commutivity parameter $\theta$ in
Yang-Mills instantons. This similarity was previously noted in \cite{wittphase,nek}, 
and we shall make the analogy more explicit in the following section. 
\para
The Manton metric on $\hat{\cal V}_{k,(N,M)}$ may be once again 
defined in terms of the overlap of zero modes. The resulting metric 
is K\"ahler and inherits a $S(U(N)\times U(M))$ holomorphic isometry from 
the surviving symmetry group \eqn{survivor}, together with a further 
$U(1)$ isometry from the rotational symmetry. However, the Manton 
metric on $\hat{\cal V}_{k,(N,M)}$ suffers from a sickness since some 
of the zero modes are (logarithmically) non-normalisable. 
This well known problem for lumps in the ${\bf CP}^{M}$ sigma-model 
\cite{ward} is not ameliorated by a finite gauge coupling 
$e^2$ as shown in \cite{batsam}. Classically this non-normalisability 
ensures that certain collective coordinates (for example, the scaling size 
of the lump) have infinite moment of 
inertia and are thus constants of the dynamics. The non-normalisability of 
modes leads to subtleties when treating these objects 
quantum mechanically.

\begin{figure}[htb]
\begin{center}
\epsfxsize=4.5in\leavevmode\epsfbox{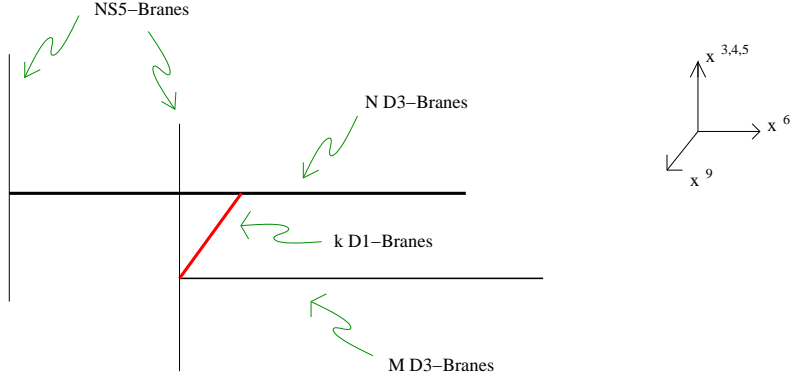}
\end{center}
\caption{The brane configuration for $U(N)$ gauge theory with 
$N+M$ hypermultiplets, and $k$ vortices.}
\label{vibfig3}
\end{figure}
\subsubsection*{Branes}

We now turn to the brane realisation of these vortices. We keep the 
same basic structure as Section 3, simply adding $M$ further 
semi-infinite D3-branes to provide the extra flavours. We choose 
to add these to the right-hand NS5-brane, so the final set-up 
is shown in Figure 3. 
\para
Once again, it is a simple matter to read off the theory on the 
$k$ D1-branes \cite{bh}. It consists of a $U(k)$ field theory, still coupled 
to the chiral multiplets $Z$ and $\psi$ as in Section 3, but now 
augmented with $M$ further chiral multiplets $\tilde{\psi}$ 
which transform in the $\bar{\bf k}$ representation of the 
gauge group. We shall write,
\be
\tilde{\psi}=\tilde{\psi}^w_{\ m}\ \ \ \ \ \ \ \ \ \ \ m=1,\ldots,k\ ;\ 
w=1,\ldots, M
\nn\ee
These fields also transform under their own $U(M)_E$ flavour 
symmetry, so the full global symmetry group of the theory is 
therefore
\be
G=SU(2)_R\times S(U(N)_D\times U(M)_E)\times U(1)_F
\nn\ee
where the overall $U(1)$ of the $U(N)_D\times U(M)_E$ flavour 
symmetry lies in the $U(k)$ gauge group. 
\para
As in Section 3, we are interested in the Higgs branch of the 
D1-brane theory, which we denote as $\hat{\cal M}_{k,(N,M)}$. This 
Higgs branch is expected to be isomorphic  
to the vortex moduli space, 
\be
\hat{\cal V}_{k,(N,M)}\cong\hat{\cal M}_{k,(N,M)}.
\nn\ee
Let us examine the Higgs branch in more detail. It is 
given by a $U(k)$ quotient of ${\bf C}^{k(N+M+k)}$, parameterised 
by $Z$, $\psi$ and $\tilde{\psi}$. The D-term moment map is
\be
D^m_{\ n}=\sum_{i=1}^N\psi^m_{\ i}\psi^{i\dagger}_{\ n}
-\sum_{w=1}^M\tilde{\psi}^{m\dagger}_{\ w}\tilde{\psi}^w_{\ n}
-[Z,Z^\dagger]^m_{\ n}-r\delta^m_{\ n}=0
\label{semid}\ee
where we have, for once,  abandoned the summation convention in order to 
highlight the ranges of the various indices. The D-term imposes 
$k^2$ real constraints  which are augmented by the restriction 
to $U(k)$ invariant coordinates. Thus the real dimension of the 
Higgs branch is
\be
{\rm dim}(\hat{\cal M}_{k,(N,M)})=2k(N+M)
\nn\ee
in agreement with the vortex moduli space. The symmetry group $G$ 
imprints itself as a holomorphic $S(U(N)\times U(M))\times U(1)$ 
isometry of the Higgs branch and thus, as before, the symmetries 
of the metric on the Higgs branch defined by the K\"ahler 
quotient construction are the same as those of the Manton 
metric. However, as in Section 3, here the agreement stops. 
In particular, the metric on $\hat{\cal M}_{k,(N,M)}$ defined by 
the K\"ahler quotient construction is finite, and sees nothing 
of the non-normalisable modes of the vortex. Given our remarks in 
the introduction, one would expect that for many supersymmetric 
problems in soliton and instanton physics, one can replace the 
Manton metric on $\hat{\cal V}_{k\,(N,M)}$ with the K\"ahler 
quotient metric on $\hat{\cal M}_{k,(N,M)}$ and in this manner 
avoid the subtleties of the non-normalisable directions. 
\para
The FI parameter of \eqn{semid} is once 
again related to the gauge coupling,
\be
r=\frac{2\pi}{e^2}
\nn\ee
so that the limit in which the semi-local vortices reduce to 
sigma model lumps is simply $r\rightarrow 0$. 
Let examine how the Higgs branch changes with $r$. For the vortex theory described in 
Section 3, the Higgs branch only exists for $r>0$. When 
$r=0$, the D-term \eqn{D} requires $\psi=0$, and 
the metric on the Higgs branch becomes the flat, singular 
metric on ${\bf C}^k/S^k$. This 
reflects the fact that the vortices of Section 2 
become point-like objects when $e^2\rightarrow \infty$. 
However, things are somewhat different for the semi-local 
vortices. In this case, the moduli space is smooth for 
$r>0$, and again develops singularities when $r=0$. These 
singularities correspond to the zero size sigma-model 
lumps. Yet, even for $r=0$,  there still exist 
solutions to the D-term equations \eqn{semid} with 
$\psi\, ,\, \tilde{\psi}\neq 0$. These correspond to the 
non-singular sigma-model lumps of finite size. Note that for 
semi-local vortices the Higgs branch defined by \eqn{semid} 
even makes sense for $r<0$. One may want to interpret this as a 
``continuation past infinite coupling'' ($e^2\rightarrow\infty$ and 
then some). 
In the following Section we shall give a different interpretation for 
the regime $r<0$.

\subsubsection*{Instantons}

The moduli space $\hat{\cal V}_{k,(N,M)}$ of $k$ vortices in $U(N)$ 
theories with $N_f=(N+M)$ flavours is again a complex submanifold 
of the moduli space of instantons. This time we must look at 
${\cal I}_{k,N_f}$ describing $k$ instantons in $U(N_f)$ 
Yang-Mills theory. The ADHM theory for these instantons 
was described in Section 4 where, obviously, we must replace 
$N$ with $N_f$. So, for example, the global symmetry group 
of the ADHM theory is,
\be
G^\prime=Spin(5)_R\times SU(N_f)_F\times SU(2)_L\times U(1)_R
\nn\ee
As in Section 4, we define the submanifold describing vortices by 
weakly gauging a symmetry or, equivalently, by the fixed point set 
of a ${\bf S}^1$ action on ${\cal I}_{k,N_f}$. The new ingredient 
here is that the $\widehat{U(1)}$ action includes a component from the 
$SU(N_f)_F$ flavour symmetry. In fact, it will prove to be simpler 
to phrase the discussion in terms of $U(1)_G\times SU(N_f)_F = U(N_f)$, 
where $U(1)_G\subset U(k)$ is the overall $U(1)$ gauge symmetry. 
To this end, consider the Cartan subalgebra of 
$U(N_f)$, 
\be
%\otimes_{i=1}^{N+M}
\prod_{i=1}^{N_f}
\ U(1)_F^{(i)}
\nn\ee
Write the associated 
charges as $Q_F^{(i)}$, where $i=1,\ldots,N_f$. In this notation, the 
charge under the overall $U(1)$ gauge symmetry is,
\be
Q_G=\sum_{i=1}^{N_f}Q_F^{(i)}
\nn\ee
The theory describing semi-local vortices can be obtained from the ADHM theory by a 
weak gauging which gives mass to all fields carrying non-vanishing charge,
\be
\widehat{Q}=Q_A+Q_R-Q_L-\sum_{i=1}^NQ_F^{(i)}+\sum_{i=N+1}^M
Q_F^{(i)}
\nn\ee
and the vortex moduli space $\hat{\cal V}_{k,(N,M)}$ is isomorphic 
to the fixed point set of the associated $\widehat{U(1)}$ action
on the instanton moduli space ${\cal I}_{k,N+M}$. In this case, 
the $\widehat{U(1)}$ action arises from a simultaneous rotation of 
the instantons in the $x^3-x^4$ plane, together with a gauge rotation 
in $U(N_f)$.

\section{Non-Commutative Vortices} 

In this Section we examine our Yang-Mills-Higgs theories defined the spatial 
non-commutative plane 
\be
[x^1,x^2]=-i\vartheta
\label{ncvort}\ee
and ask how this deformation affects the moduli space of vortices. Various aspects 
of non-commutative vortices in 
the abelian Higgs model have been considered in \cite{indians,blp,loz,mencvort}. 
Before recalling the results of these papers, let us 
start by regaling ourselves with the beautiful tale of non-commutative 
instantons. We have already covered this is Section 4, but a good story is 
always worth retelling. It was shown by Nekrasov and Schwarz \cite{ns} 
that a non-commutative deformation of ${\bf R}^4$ 
as given, for example, in \eqn{noncom} induces a FI parameter $r=4\theta$ in 
the ADHM instanton gauge theory. This FI term resolves the singularities of 
the instanton moduli space in the manner described previously by Nakajima \cite{nak}. 
One may expect that a similar phenomenon occurs in our vortex theory. However, 
we have seen that the job of resolving the singularities on the vortex 
moduli space is already adequately performed by the gauge coupling since 
the FI parameter is $r=2\pi/e^2$. So what role could the non-commutative deformation 
\eqn{ncvort} play? To avoid undue suspense, we shall first reveal the answer, followed 
by a derivation, and then an analysis of the consequences. 
We shall show that the effect of a non-commutative background is to change the FI 
parameter of the vortex theory to 
\be
r=2\pi\left(\frac{1}{e^2}+\vartheta\zeta\right)
\nn\ee
To see this, we return once more to the brane picture of Section 3. We want 
now to deform the D3-brane dynamics so that at low energies it is described 
by the Yang-Mills-Higgs theory \eqn{lag} defined on the 
non-commutative plane \eqn{ncvort}. 
The string theory background 
that achieves this feat is well known \cite{sw}: a background NS NS 
$B$ field is added, with components $B_{12}\neq 0$.  We wish to 
understand the effect of this $B$ field on the dynamics of the D1-strings. 
\para
In fact, a very similar situation was analysed by Hashimoto and Hashimoto 
in \cite{hash2}. These authors considered the situation of a D-string 
suspended between two D3-branes in a background $B$ field, a set-up which 
describes  
a monopole in non-commutative Yang-Mills. The basic physics is very 
simple to describe. The background $B$ field may be absorbed into the 
D3-brane as a magnetic flux $F_{12}$. The end of the D1-brane acts as a 
magnetic source in the D3-brane, and therefore experiences a force due 
to $F_{12}$. To see how force acts, recall that our D3-brane lies in the 
$0126$ directions and note that $F_{12}={}^\star F_{06}$. The magnetic end of the 
D-string therefore feels the same force as an electric charge in a 
background electric field $F_{06}$. In other words, the string end moves 
in the $x^6$ direction. This displacement continues until 
the force due to the $B$ field is canceled 
by the excess tension of the D1-brane. In our case, one end of the D1-brane 
is attached to the NS5-brane, and cannot move in the $x^6$ direction. The final 
configuration is therefore given by the tilted D-strings, as shown in 
Figure 4. An analysis of the supersymmetry generators was performed 
in \cite{hash2} which, translated to the present set-up, 
reveals that these tilted D1-branes continue to preserve four supercharges. 
\begin{figure}[htb]
\begin{center}
\epsfxsize=4.5in\leavevmode\epsfbox{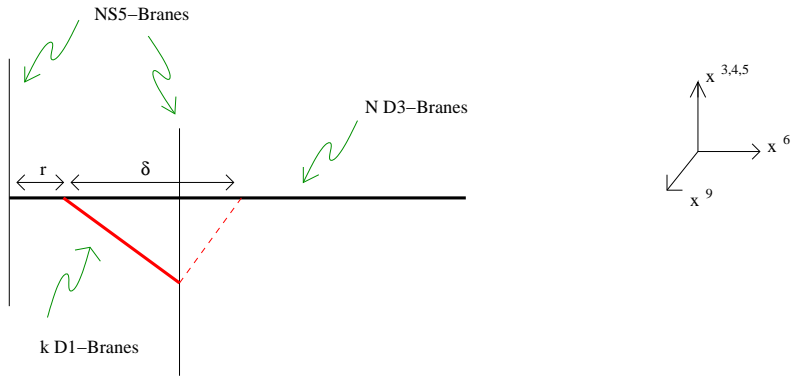}
\end{center}
\caption{The brane configuration for vortices on a non-commutative 
background. A NS-NS B field lies in the $x^1-x^2$ directions, inducing 
non-commutivity on the D3-brane worldvolume and causing the D1-brane 
to tilt.}
\label{vibfig4}
\end{figure}
\para
The effect of this tilt on the theory on the D1-branes is to change the 
FI parameter $r$, which is now given (up to a normalisation of $1/g_s$ - see 
\eqn{rel2}) by the distance between the end of 
the D1-brane and the left-hand NS5-brane. The displacement $\delta$ of the D1-brane 
from its original position was calculated in \cite{hash2}, and is given by
\be
\delta=-\frac{\vartheta\,\Delta x^9}{2\pi l_s^2}
\nn\ee
With the parameters $e^2$ and $\zeta$ still defined in terms of the distances 
$\Delta x^6$ and $\Delta x^9$ between the NS5-branes as in \eqn{rel1}, 
we find the promised result,
\be
r=2\pi\left(\frac{1}{e^2}+\vartheta\zeta\right).
\label{ohyes}\ee
We now turn to studying some simple consequences of this equation. Consider 
first the vortices of Section 2, in which we have a $U(N)$ gauge group 
with $N$ hypermultiplets. The most striking feature is that the vortex 
moduli space only exists for $r\geq0$ or, alternatively, for $\vartheta$ 
above the critical value $\vartheta_c$
\be
\vartheta>\vartheta_c = -1/e^2\zeta
\nn\ee
Moreover, at the critical value $\vartheta=\vartheta_c$, 
the moduli space becomes singular. For example, in the abelian case $N=1$, 
we have ${\cal V}_{k,1}\cong {\bf C}^k/S_k$, endowed 
with the flat, singular metric at $\vartheta=\vartheta_c$. 
In fact neither of these results are new: 
both were 
previously derived in the abelian $N=1$ theories by Bak, Lee and Park 
by an explicit study of the solutions 
to the non-commutative Bogomoln'yi equations \cite{blp} (see also 
the phase diagram in \cite{mencvort}). It is pleasing  to see these 
properties reproduced from our D1-brane theory. 
From equation \eqn{ohyes} it 
is also clear that vortex moduli space $\vkn$ is non-singular in the 
$e^2\rightarrow\infty$ limit provided $\vartheta >0$. This point was previously 
made in \cite{mencvort} and the Manton metric on ${\cal V}_{k,1}$ 
was explicitly calculated in this limit. Once again, the Manton metric 
differs from the metric induced on the Higgs branch by the K\"ahler 
quotient construction. For example, in the case of two vortices, the 
non-commutative metric in the $e^2\rightarrow \infty$ limit takes 
the form of a cone \eqn{cone}, now with the conformal factor given by
\be
f^2(\sigma)=\coth(\sigma^2/2)-\frac{\sigma^2}{2\sinh^2(\sigma^2/2)}\approx
1-2\sigma^2e^{-\sigma^2}+\ldots
\nn\ee
which coincides with neither \eqn{2vort} nor \eqn{man2vort}.
\para
We now turn to the consequences of \eqn{ohyes} for semi-local vortices. 
Firstly, note that if we set $e^2\rightarrow\infty$ so that we 
are studying the moduli space of sigma-model lumps, then the non-commutivity 
parameter $\vartheta$ resolves the singularities just as in the case 
of Yang-Mills instantons. Similar observations were made from a field 
theory perspective in \cite{leeleeyang}. However, for semi-local 
vortices the moduli space defined by the D-term \eqn{semid} continues to make 
sense for $r<0$ or, alternatively, for $\vartheta<\vartheta_c$. Moreover, the 
topology of the moduli space differs for $r>0$ and $r<0$. Thus, as we decrease 
the non-commutivity parameter $\vartheta$ past its critical value of $-1/e^2\zeta$, 
the moduli space of non-commutative vortices undergoes a topology changing 
transition. A familiar example of this occurs for a single $k=1$ 
semi-local vortex in a $U(2)$ gauge theory with 4 hypermultiplets. In this 
case the moduli space is related to the Calabi-Yau 3-fold known to string 
theorists simply as The Conifold. For $r=0$, it is the singular space 
defined by the complex equation $xy-wz=0$. The FI parameter resolves the 
singularity and we have,
\be
\hat{\cal V}_{1,(2,2)}\cong O(-1)\times O(-1)\rightarrow {\bf CP}^1
\nn\ee
As we decrease the non-commutivity parameter $\vartheta$ 
to pass from $r>0$ to $r<0$, the 
moduli space of vortices undergoes a flop transition. For other values of $N$ and 
$M$, the topology change is more dramatic and the moduli space of vortices 
may have different Betti numbers for $r>0$ and $r<0$. 
\para
Finally, note that for $\vartheta<\vartheta_c$ the equation \eqn{ohyes} implies a 
duality between 
different non-commutative gauge theories in the sense that they share 
the same moduli spaces of vortices. 
Specifically, consider the $U(N)$ theory with $N_f=(N+M)$ 
flavours. The moduli space of $k$ vortices 
$\hat{\cal V}_{k,(N,M)}$ is defined by the moment map \eqn{semid}  
with FI parameter $r$ given by \eqn{ohyes}. For $\vartheta < -1/e^2\zeta$ 
we have $r<0$. From the D-term \eqn{semid} we see that the 
moduli space of vortices in this theory coincides with 
the moduli space of vortices in a $U(M)$ gauge theory, also with 
$N_f=(N+M)$ flavours. 
(We must also perform a parity transformation in going from the 
$U(N)$ theory to the $U(M)$ theory, so that $Z\rightarrow Z^\dagger$).
Denote the gauge coupling of the $U(M)$ 
theory as $e^{\prime 2}$, and the non-commutivity parameter as $\vartheta^\prime$. 
Then for the moduli spaces of vortices to coincide, we simply require
\be
\frac{1}{e^{\prime2}}+\vartheta^\prime\zeta = -\frac{1}{e^2}-\vartheta\zeta
\nn\ee 
This duality is, like many, reminiscent of the $(N_c=N)\rightarrow (N_f-N_c=M)$ 
duality of Seiberg.

\section{Summary and Discussion}

Let us begin this ending with a summary of our results. We have studied vortices in 
$U(N)$ Yang-Mills theories with $N_f\geq N$ flavours. These theories have a FI 
parameter $\zeta$ which ensures the gauge group is completely broken in the 
vacuum. The gauge coupling parameter is $e^2$ and the theories may be defined on 
the non-commutative plane with $[x^1,x^2]=-i\vartheta$. 
\para
We have shown that 
the moduli space of charge $k$ vortices in this theory 
is described by the Higgs branch of a 
$U(k)$ gauge theory with four supercharges, coupled to $N$ chiral multiplets 
$\psi$ transforming in the $k$ representation of the gauge group, $N_f-N$ 
chiral multiplets $\tilde{\psi}$ transforming in the $\bar{k}$ representation, and a single 
chiral multiplet $Z$ transforming in the adjoint representation. This Higgs 
branch is defined by a $U(k)$ symplectic quotient of ${\bf C}^{k(k+N_f)}$ 
with moment map,
\be
[Z^\dagger,Z]+\psi\psi^\dagger-\tilde{\psi}^\dagger\tilde{\psi} = r
\label{theend}\ee
where the level of the moment map $r$ is a FI parameter defined by
\be
r=2\pi\left(\frac{1}{e^2}+\vartheta\zeta\right)
\nn\ee
We further showed that the vortex moduli space 
may be constructed as a complex submanifold 
(or, for $N_f=N$, a special Lagrangian submanifold) of the moduli space 
of $k$ instantons in non-commutative $U(N_f)$ Yang-Mills theory.
\para
Relationships between the instanton and vortex equations have been noted 
in the past. In particular, a reduction of instantons in $SU(2)$ Yang-Mills 
on ${\bf R}^2\times {\bf S}^2$ gives rise to the $U(1)$ vortex equations 
\cite{fm}. While this relationship appears to share several characteristics 
of our correspondence, it differs in many important 
details. It would be interesting 
to elucidate the  connections between these two approaches.
\para
As we discussed in detail, the Manton metric on the moduli space of vortices  does not 
coincide with the metric induced on the Higgs branch \eqn{theend} by the K\"ahler 
quotient construction. Indeed, all our experience with gauge theories with four 
supercharges suggests that it is too ambitious an enterprise to determine 
the metric on the moduli space of vortices. Other, more topological, questions 
can be asked with greater success and the construction \eqn{theend} provides the 
answers. As stressed in the introduction, these topological questions include 
certain quantum correlation functions in supersymmetric gauge theories.
\para
Given the relationship described in Section 4 between the moduli space 
of vortices and the moduli space of Yang-Mills instantons, one may expect 
quantitative agreement between topological correlation functions of two 
and four dimensional gauge theories. These would 
receive non-perturbative instanton corrections in four dimensions, 
and vortex corrections in two dimensions. 
Indeed, it is well known 
that ${\cal N}=1$ $SU(M+1)$ super-Yang-Mills in four dimensions 
shares many features with the ${\cal N}=(2,2)$ ${\bf CP}^{M}$ sigma-model 
in two dimensions including, most pertinently, its  
low-energy effective action \cite{vy,cpn}. A second, more quantitative, example 
was given in \cite{dorey} where ${\cal N}=2$ theories in four dimensions 
were shown to have a spectrum of monopoles that coincides with the spectrum of kinks 
in certain ${\cal N}=(2,2)$ theories in two dimensions. Both 
monopole and kink masses receive identical corrections from non-perturbative 
effects, namely Yang-Mills instantons and vortices respectively. 
It seems likely that the semi-classical reason for these agreements can be traced 
to the relationship between the vortex and instanton moduli spaces described 
here.

\section*{Acknowledgments}
We would like to thank Fred Goldhaber, Carlos Nunez, Martin Schnabl, Ashoke Sen, 
Wati Taylor and Jan Troost for taking the time to talk to us. 
A.~H. is supported in part by the Reed Fund Award and a DOE OJI Award.
D.~T. is supported by a Pappalardo fellowship, and is very grateful 
to the Pappalardo family for their generosity.
This work was also supported in part 
by funds provided by the U.S. Department of Energy (D.O.E.) under 
cooperative research agreement \#DF-FC02-94ER40818.


\begin{thebibliography}{99}

\small
\parskip=0pt plus 2pt

\bibitem{manton} N.~S.~Manton, ``{\em A Remark On The Scattering Of BPS Monopoles}'', 
Phys.\ Lett.\ B {\bf 110}, 54 (1982).
%%CITATION = PHLTA,B110,54;%%
\bibitem{adhm} M.~F.~Atiyah, N.~J.~Hitchin, V.~G.~Drinfeld and Y.~I.~Manin,
``{\em Construction Of Instantons}'',
Phys.\ Lett.\ A {\bf 65}, 185 (1978).
%%CITATION = PHLTA,A65,185;%%
\bibitem{nahm} W. Nahm, 
``{\em The Construction Of All Selfdual Multi-Monopoles By The ADHM Method}'', 
in Monopoles in Quantum Field Theory, eds. N.S. Craigie, P. Goddard and W. Nahm.
\bibitem{ah} M.~F.~Atiyah and N.~J.~Hitchin,
``{\em The Geometry And Dynamics Of Magnetic Monopoles}'', 
Princeton University Press. 
\bibitem{samols} T.~M.~Samols,
``{\em Vortex Scattering}''
Commun.\ Math.\ Phys.\  {\bf 145}, 149 (1992).
%%CITATION = CMPHA,145,149;%%
\bibitem{manspe} N.~S.~Manton and J.~M.~Speight,
``{\em Asymptotic interactions of critically coupled vortices}''
Commun.\ Math.\ Phys.\  {\bf 236}, 535 (2003)
[arXiv:hep-th/0205307].
%%CITATION = HEP-TH 0205307;%%
\bibitem{witten81} E.~Witten,
``{\em Constraints On Supersymmetry Breaking}''
Nucl.\ Phys.\ B {\bf 202}, 253 (1982).
%%CITATION = NUPHA,B202,253;%%
\bibitem{boundstate} For the general theory with specific application to monopoles, 
see the lecture notes by  \vspace{0.1cm} \\  
J.~A.~Harvey, ``{\em Magnetic monopoles, duality, and supersymmetry}''
arXiv:hep th/9603086.
%%CITATION = HEP-TH 9603086;%%
\bibitem{instanton} For recent applications to ${\cal N}=2$ 
gauge theories in four dimensions, see \vspace{0.1cm} \\
N.~A.~Nekrasov,
``{\em Seiberg Witten prepotential from instanton counting}''
arXiv:hep-th/0206161. \\ 
%%CITATION = HEP-TH 0206161;%% 
T.~J.~Hollowood,
``{\em Testing Seiberg Witten theory to all orders in the instanton expansion}''
Nucl.\ Phys.\ B {\bf 639}, 66 (2002)
[arXiv:hep-th/0202197]. \\
%%CITATION = HEP-TH 0202197;%%
R.~Flume and R.~Poghossian,
``{\em An algorithm for the microscopic evaluation of the coefficients of the  
Seiberg-Witten prepotential}''
arXiv:hep-th/0208176. 
%%CITATION = HEP-TH 0208176;%%
%U.~Bruzzo, F.~Fucito, J.~F.~Morales and A.~Tanzini,
%``{\em Multi-instanton calculus and equivariant cohomology}''
%arXiv:hep-th/0211108.
%%CITATION = HEP-TH 0211108;%%
\bibitem{instanton2} An elegant application in three dimensional 
gauge theories can be found in \vspace{0.1cm}\\ 
N.~Dorey, V.~V.~Khoze and M.~P.~Mattis,
``{\em, Multi-instantons, three-dimensional gauge theory, and the  
Gauss-Bonnet-Chern theorem}''
Nucl.\ Phys.\ B {\bf 502}, 94 (1997)
[arXiv:hep-th/9704197].
%%CITATION = HEP-TH 9704197;%%
\bibitem{bog} E. B. Bogomol'nyi, ``{\em The stability of classical 
solutions}'', Soviet J. Nucl. Phys. {\bf 24} (1976) 449.
\bibitem{erick} E.~J.~Weinberg,
``{\em Multivortex Solutions Of The Ginzburg-Landau Equations}'', 
Phys.\ Rev.\ D {\bf 19}, 3008 (1979). \\ 
%%CITATION = PHRVA,D19,3008;%%
E.~J.~Weinberg,
``{\em Index Calculations For The Fermion - Vortex System}''
Phys.\ Rev.\ D {\bf 24}, 2669 (1981).
%%CITATION = PHRVA,D24,2669;%%
\bibitem{taubes} A.~Jaffe and C.~Taubes,
``{\em Vortices And Monopoles. Structure Of Static Gauge Theories}'', 
Birkhaeuser (1980).
\bibitem{tong} D.~Tong,
``{\em NS5-branes, T-duality and worldsheet instantons}''
JHEP {\bf 0207}, 013 (2002)
[arXiv:hep-th/0204186].
%%CITATION = HEP-TH 0204186;%%
\bibitem{spe} 
J.~M.~Speight,
``{\em Static intervortex forces}''
Phys.\ Rev.\ D {\bf 55}, 3830 (1997) [arXiv:hep-th/9603155].
%%CITATION = PHRVA,D55,3830;%%
\bibitem{hw} A.~Hanany and E.~Witten,
``{\em Type IIB superstrings, BPS monopoles, and three-dimensional gauge  dynamics}''
Nucl.\ Phys.\ B {\bf 492}, 152 (1997)
[arXiv:hep-th/9611230].
%%CITATION = HEP-TH 9611230;%%
%\cite{Brodie:1997sz}
\bibitem{bh}
J.~H.~Brodie and A.~Hanany,
``{\em Type IIA superstrings, chiral symmetry, and N = 1 4D gauge theory  dualities}''
Nucl.\ Phys.\ B {\bf 506}, 157 (1997)
[arXiv:hep-th/9704043].
%%CITATION = HEP-TH 9704043;%%
\bibitem{hh} A.~Hanany and K.~Hori,
``{\em Branes and N = 2 theories in two dimensions}''
Nucl.\ Phys.\ B {\bf 513}, 119 (1998)
[arXiv:hep-th/9707192].
%%CITATION = HEP-TH 9707192;%%
\bibitem{lee} C.~Kim, K.~Lee and S-H.~Yi,
``{\em Tales of D0 on D6 branes: Matrix mechanics of identical particles}''
Phys.\ Lett.\ B {\bf 543}, 107 (2002)
[arXiv:hep-th/0204109].
%%CITATION = HEP-TH 0204109;%%
\bibitem{poly} A.~P.~Polychronakos,
``{\em Quantum Hall states as matrix Chern-Simons theory}''
JHEP {\bf 0104}, 011 (2001)
[arXiv:hep-th/0103013].
%%CITATION = HEP-TH 0103013;%%
\bibitem{adhmgauge} E.~Witten,
``{\em Small Instantons in String Theory}''
Nucl.\ Phys.\ B {\bf 460}, 541 (1996)
[arXiv:hep-th/9511030]. \\
%%CITATION = HEP-TH 9511030;%%
M.~R.~Douglas,
``{\em Gauge Fields and D-branes}''
J.\ Geom.\ Phys.\  {\bf 28}, 255 (1998)
[arXiv:hep-th/9604198].
%%CITATION = HEP-TH 9604198;%%
\bibitem{ns} N.~Nekrasov and A.~Schwarz,
``{\em Instantons on noncommutative $R^4$ and (2,0) superconformal six  dimensional theory}''
Commun.\ Math.\ Phys.\  {\bf 198}, 689 (1998)
[arXiv:hep-th/9802068].
%%CITATION = HEP-TH 9802068;%%
\bibitem{nak} H. Nakajima, ``{\em Lectures on Hilbert schemes of points on 
surfaces}'', AMS Lecture Series, Vol 18.
\bibitem{pot} Potentials on monopole moduli spaces were introduced in the 
study of dyon dynamics
\vspace{0.1cm} \\ 
 D.~Tong,
``{\em A note on 1/4-BPS states}''
Phys.\ Lett.\ B {\bf 460}, 295 (1999)
[arXiv:hep-th/9902005]. \\
%%CITATION = HEP-TH 9902005;%%
D.~Bak, C.~Lee, K.~Lee and P.~Yi,
``{\em Low energy dynamics for 1/4 BPS dyons}''
Phys.\ Rev.\ D {\bf 61}, 025001 (2000)
[arXiv:hep-th/9906119].
%%CITATION = HEP-TH 9906119;%%
\bibitem{potinst} Discussions of potentials on instanton moduli spaces 
have been given in  \vspace{0.1cm} \\ 
N.~D.~Lambert and D.~Tong,
``{\em Dyonic instantons in five-dimensional gauge theories}''
Phys.\ Lett.\ B {\bf 462}, 89 (1999)
[arXiv:hep-th/9907014]. \\
%%CITATION = HEP-TH 9907014;%%
N.~Dorey, T.~J.~Hollowood, V.~V.~Khoze and M.~P.~Mattis,
``{\em The calculus of many instantons}''
Phys.\ Rept.\  {\bf 371}, 231 (2002)
[arXiv:hep-th/0206063].  
%%CITATION = HEP-TH 0206063;%%
\bibitem{leeyi} K.~~Lee and P.~Yi,
``{\em Quantum spectrum of instanton solitons in five dimensional  
noncommutative U(N) theories}''
Phys.\ Rev.\ D {\bf 61}, 125015 (2000)
[arXiv:hep-th/9911186].
%%CITATION = HEP-TH 9911186;%%
\bibitem{ksme} K.~Lee, D.~Tong and S.~Yi,
``{\em The moduli space of two U(1) instantons on noncommutative $R^4$ and 
$R^3\times S^1$}'' Phys.\ Rev.\ D {\bf 63}, 065017 (2001)
[arXiv:hep-th/0008092].
%%CITATION = HEP-TH 0008092;%%
\bibitem{vetal} M.~Bershadsky, C.~Vafa and V.~Sadov,
``{\em D-Branes and Topological Field Theories}''
Nucl.\ Phys.\ B {\bf 463}, 420 (1996)
[arXiv:hep-th/9511222]. \vspace{0.1cm} \\ 
%%CITATION = HEP-TH 9511222;%%
For lectures on wrapped branes, see \vspace{0.1cm} \\ 
J.~P.~Gauntlett,
``{\em Branes, calibrations and supergravity}''
arXiv:hep-th/0305074.
%%CITATION = HEP-TH 0305074;%%
\bibitem{ht} C.~M.~Hull and P.~K.~Townsend,
``{\em Unity of superstring dualities}''
Nucl.\ Phys.\ B {\bf 438}, 109 (1995)
[arXiv:hep-th/9410167].
%%CITATION = HEP-TH 9410167;%%
%\bibitem{who} A.~Karch, D.~Lust and D.~J.~Smith,
%``{\em Equivalence of geometric engineering and Hanany-Witten via 
%fractional  branes}'', 
%Nucl.\ Phys.\ B {\bf 533}, 348 (1998)
%[arXiv:hep-th/9803232].
%%CITATION = HEP-TH 9803232;%%
\bibitem{wittphase} E.~Witten,
``{\em Phases of N = 2 theories in two dimensions}''
Nucl.\ Phys.\ B {\bf 403}, 159 (1993)
[arXiv:hep-th/9301042].
%%CITATION = HEP-TH 9301042;%%
\bibitem{vash} A.~Achucarro and T.~Vachaspati,
``{\em Semilocal and electroweak strings}''
Phys.\ Rept.\  {\bf 327}, 347 (2000)
[Phys.\ Rept.\  {\bf 327}, 427 (2000)]
[arXiv:hep-ph/9904229].
%%CITATION = HEP-PH 9904229;%%
\bibitem{gibbons} G.~W.~Gibbons, M.~E.~Ortiz, F.~Ruiz Ruiz and T.~M.~Samols,
``{\em Semilocal strings and monopoles}''
Nucl.\ Phys.\ B {\bf 385}, 127 (1992)
[arXiv:hep-th/9203023].
%%CITATION = HEP-TH 9203023;%%
\bibitem{tongdw} D.~Tong,
``{\em The moduli space of BPS domain walls}''
Phys.\ Rev.\ D {\bf 66}, 025013 (2002)
[arXiv:hep-th/0202012].
%%CITATION = HEP-TH 0202012;%%
\bibitem{bert} 
B.~J.~Schroers,
``{\em The Spectrum of Bogomol'nyi Solitons in Gauged Linear Sigma Models}''
Nucl.\ Phys.\ B {\bf 475}, 440 (1996)
[arXiv:hep-th/9603101].
%%CITATION = HEP-TH 9603101;%%
\bibitem{nek} A.~Losev, N.~Nekrasov and S.~L.~Shatashvili,
``{\em The freckled instantons}''
arXiv:hep-th/9908204.
%%CITATION = HEP-TH 9908204;%%
\bibitem{ward} R.~S.~Ward,
``{\em Slowly Moving Lumps In The $CP^1$ Model In (2+1)-Dimensions}''
Phys.\ Lett.\ B {\bf 158}, 424 (1985).
%%CITATION = PHLTA,B158,424;%%
\bibitem{batsam} R.~A.~Leese and T.~M.~Samols,
``{\em Interaction of semilocal vortices}'', 
Nucl.\ Phys.\ B {\bf 396}, 639 (1993).
%%CITATION = NUPHA,B396,639;%%
\bibitem{sw} 
N.~Seiberg and E.~Witten,
``{\em String theory and noncommutative geometry}''
JHEP {\bf 9909}, 032 (1999)
[arXiv:hep-th/9908142].
%%CITATION = HEP-TH 9908142;%%
\bibitem{hash2} A.~Hashimoto and K.~Hashimoto,
``{\em Monopoles and dyons in non-commutative geometry}'', 
JHEP {\bf 9911}, 005 (1999)
[arXiv:hep-th/9909202].
%%CITATION = HEP-TH 9909202;%%
\bibitem{indians}
D.~P.~Jatkar, G.~Mandal and S.~R.~Wadia,
``{\em Nielsen-Olesen vortices in noncommutative Abelian Higgs model}''
JHEP {\bf 0009}, 018 (2000)
[arXiv:hep-th/0007078].
%%CITATION = HEP-TH 0007078;%%
\bibitem{blp} D.~Bak, K.~Lee and J.~H.~Park,
``{\em Noncommutative vortex solitons}''
Phys.\ Rev.\ D {\bf 63}, 125010 (2001)
[arXiv:hep-th/0011099].
%%CITATION = HEP-TH 0011099;%%
\bibitem{loz} G.~S.~Lozano, E.~F.~Moreno and F.~A.~Schaposnik,
``{\em Nielsen-Olesen vortices in noncommutative space}'',
Phys.\ Lett.\ B {\bf 504}, 117 (2001)
[arXiv:hep-th/0011205].
%%CITATION = HEP-TH 0011205;%%
\bibitem{mencvort}  D.~Tong,
``{\em The moduli space of noncommutative vortices}'', 
J.\ Math.\ Phys.\  {\bf 44}, no. 8 (2003)
[arXiv:hep-th/0210010].
%%CITATION = HEP-TH 0210010;%%
\bibitem{leeleeyang}
B.~H.~Lee, K.~M.~Lee and H.~S.~Yang,
``{\em The $CP^n$ model on noncommutative plane}''
Phys.\ Lett.\ B {\bf 498}, 277 (2001)
[arXiv:hep-th/0007140]. \\ 
%%CITATION = HEP-TH 0007140;%%
K.~Furuta, T.~Inami, H.~Nakajima and M.~Yamamoto,
``{\em Low-energy dynamics of noncommutative $CP^1$ solitons in 2+1 dimensions}''
Phys.\ Lett.\ B {\bf 537}, 165 (2002)
[arXiv:hep-th/0203125]. \\ 
%%CITATION = HEP-TH 0203125;%%
J.~Murugan and R.~Adams,
``{\em Comments on noncommutative sigma models}''
JHEP {\bf 0212}, 073 (2002)
[arXiv:hep-th/0211171].
%%CITATION = HEP-TH 0211171;%%
\bibitem{fm} 
P.~Forgacs and N.~S.~Manton,
``{\em Space-Time Symmetries In Gauge Theories}''
Commun.\ Math.\ Phys.\  {\bf 72}, 15 (1980).
%%CITATION = CMPHA,72,15;%%
\bibitem{vy} G.~Veneziano and S.~Yankielowicz,
``{\em An Effective Lagrangian For The Pure N=1 Supersymmetric Yang-Mills Theory}''
Phys.\ Lett.\ B {\bf 113}, 231 (1982).
%%CITATION = PHLTA,B113,231;%%
\bibitem{cpn}  A.~D'Adda, A.~C.~Davis, P.~Di Vecchia and P.~Salomonson,
``{\em An Effective Action For The Supersymmetric $CP^{(N-1)}$ Model}''
Nucl.\ Phys.\ B {\bf 222}, 45 (1983).
%%CITATION = NUPHA,B222,45;%%
\bibitem{dorey} N.~Dorey,
``{\em The BPS spectra of two-dimensional supersymmetric gauge theories with  twisted 
mass terms}'', JHEP {\bf 9811}, 005 (1998)
[arXiv:hep-th/9806056]. \\
%%CITATION = HEP-TH 9806056;%%
N.~Dorey, T.~J.~Hollowood and D.~Tong,
``{\em The BPS spectra of gauge theories in two and four dimensions}''
JHEP {\bf 9905}, 006 (1999)
[arXiv:hep-th/9902134].
%%CITATION = HEP-TH 9902134;%%


\end{thebibliography}
\end{document}